\documentclass{aa}

\usepackage{graphicx}
\usepackage{amsmath}
\usepackage{subfigure}
\usepackage{kotex}
\usepackage{soul,color}
\usepackage{txfonts}
\usepackage{threeparttable}
\usepackage{bigstrut}
\usepackage{ulem}
\usepackage{booktabs}
\usepackage{multirow}
\usepackage[switch]{lineno}

\newcommand{\rph}{\ensuremath{r_{\mathrm{ph}}}}

\newcommand{\Ftdi}{\ensuremath{F_{\mathrm{TD,\infty}}}}

\newcommand{\Tbbi}{\ensuremath{T_{\mathrm{bb,\infty}}}}

\newcommand{\Finf}{\ensuremath{F_{\infty}}}
\newcommand{\fc}{\ensuremath{f_{\mathrm{c}}}}

\newcommand{\fluxunit}{\ensuremath{\mathrm{erg\,cm^{-2}\,s^{-1}}}}
\newcommand{\msun}{\ensuremath{M_{\odot}}}

\newcommand\be{\begin{eqnarray}}
\newcommand\ee{\end{eqnarray}}
\newcommand\dis{\displaystyle}

\begin{document}

\title{Measuring Masses and Radii of Neutron Stars in Low-Mass X-ray Binaries: Effects of Atmospheric Composition and Touchdown Radius}

\author{Myungkuk Kim\inst{\ref{inst1}} 
\and Young-Min Kim\inst{\ref{inst1}} 
\and Kwang Hyun Sung\inst{\ref{inst1}} 
\and Chang-Hwan Lee\inst{\ref{inst2}} 
\and Kyujin Kwak\inst{\ref{inst1}} }

\institute{Department of Physics, Ulsan National Institute of Science and Technology, Ulsan 44919, Korea\\
\email{myungkkim@unist.ac.kr,ymkim715@unist.ac.kr,kkwak@unist.ac.kr} 
\label{inst1}
\and
Department of Physics, Pusan National University, Busan 46241, Korea\\
\label{inst2}}

\date{Received ; accepted }

\abstract
{X-ray bursts (XRBs) are energetic explosive events which have been observed in  low-mass X-ray binaries (LMXBs). Some Type-I XRBs show photospheric radius expansion (PRE) and these PRE XRBs are used to simultaneously estimate the mass and the radius of a neutron star in LMXB.}
{The mass and radius estimation depends on a few model parameters most of which are still uncertain. Among them, we focus on the effects of the chemical composition of the photosphere which determines the opacity during the PRE phase and the touchdown radius which can be larger than the neutron star radius. We investigate how these two model parameters affect the mass and radius estimation in a systematic way and whether there is any statistical trend for these two parameters including a correlation between them.}
{We use both a Monte Carlo (MC) sampling and a Bayesian analysis to find the effects of the photospheric composition and the touchdown radius. We apply these two methods to six LMXBs that show PRE XRBs. In both methods, we solve the Eddington flux equation and the apparent angular area equation both of which include the correction terms. For the MC sampling, we have developed an iterative method in order to solve these two equations more efficiently.}   
{We confirm that the effects of the photospheric composition and the touchdown radius are similar in the statistical and analytical estimation of mass and radius even when the correction terms are considered. Furthermore, in all of the six sources, we find that a H-poor photosphere and a large touchdown radius are favored statistically regardless of the statistical method.
Our Bayesian analysis also hints that touchdown can occur farther from the neutron star surface when the photosphere is more H-poor. This correlation could be qualitatively understood  with the Eddington flux equation. We propose a physical explanation for this correlation between the photospheric composition and the touchdown radius.
Our results show that when accounting for the uncertainties of the photospheric composition and the touchdown radius, most likely radii of the neutron stars in these six LMXBs are less than $12.5$~km, which is similar to the bounds for the neutron star radius placed with the tidal deformability measured from the gravitational wave signal.}
{}


\titlerunning{Neutron Stars in LMXBs}
\authorrunning{Kim et al.}

\maketitle

\section{Introduction}\label{s01}

Low-mass X-ray binaries (LMXBs) are 
the nesting systems in which thermonuclear (Type I) X-ray bursts (XRBs) are found. 
When a powerful XRB occurs, the maximum X-ray luminosity can reach 
the local Eddington limit ($\ensuremath{L_{\mathrm{Edd}}} \sim 10^{38}$ erg/s) 
and it is possible that the photospheric layer is lifted off from 
the surface of the neutron star due to strong radiation pressure. 
When this happens, the photospheric radius 
can be much larger than the stellar radius of the neutron star. 
After the photospheric radius reaches its maximum, 
it returns to its original position that is usually assumed to be the surface of the neutron star. 
This phenomenon is called photospheric radius expansion (PRE) and 
about $20\%$ of all observed XRBs show the PRE feature \citep{Galloway:2006eq}. 
Although the observed flux during the PRE stage is approximately constant because it is close to 
the local Eddington flux, it actually continues to increase, reaches a peak
within a few seconds and then deceases exponentially \citep{Guver:2008gc}. This happens because the temperature 
of the photosphere increases (decreases) while the size of the photosphere (i.e., photospheric radius) 
decreases (increases). At the moment that the observed flux reaches the peak, the temperature of the 
photosphere, which can be measured from the spectral analyses, reaches its maximum. Then, the 
photospheric radius shrinks back to the minimum value which is usually assumed to be the size of the 
neutron star. This moment is called "touchdown". 

\begin{table*}[tp]
\centering
\caption{Observational properties of six LMXBs that show PRE XRBs. All values are obtained from \citet{Ozel_2016}.  \label{ta01}}
\begin{tabular}{lcccc}
\hline \hline
Source & App. angular area & Touchdown Flux & Spin Freq.\tablefootmark{a} & Distance\tablefootmark{a} \bigstrut\\
 & $\rm (km/10~kpc)^2$ & $(10^{-8}~{\fluxunit})$ & (Hz) & (kpc) \bigstrut\\
 \hline
 4U 1820--30 & 89.9$\pm$15.9 & 5.98$\pm$0.66 & ... & 7.6$\pm$0.4 (4) \bigstrut\\
 &&&&8.4$\pm$0.6 (5-6)\bigstrut\\
 SAX J1748.9--2021 & 89.7$\pm$9.6 & 4.03$\pm$0.54& 410 (1) & 8.2$\pm$0.6 (4, 5, 7)\bigstrut\\
EXO 1745--248 & 117.8$\pm$19.9 & 6.69$\pm$0.74& ... & 6.3$\pm$0.63\tablefootmark{b} (8-9)  \bigstrut\\
KS 1731--260 & 96.0$\pm$7.9 & 4.71$\pm$0.52 & 524 (2) & 7-9\tablefootmark{c} (10)\bigstrut\\
4U 1724--207 & 113.8$\pm$15.4 & 5.29$\pm$0.58 & ... & 7.4$\pm$0.5  \bigstrut\\
4U 1608--52 & 314$\pm$44.3 & 18.5$\pm$2.0 & 620 (3) & 4.0$\pm$2.0, $D_{\rm cutoff} > 3.9$\tablefootmark{d} \bigstrut\\
\hline
\end{tabular}
\vspace{0.1in}
\tablefoot{\\
The observational values, distance to target ($D$), apparent angular area ($A$), and touchdown flux (${\Ftdi}$), 
for the six targets are taken from Table 1 of \citet{Ozel_2016} except the distance to 4U~1608--52. Normal distributions for $D$, $A$, and ${\Ftdi}$ are used for both our MC sampling and Bayesian analysis. Exceptions for the $D$ distribution  are given below. \\
\tablefoottext{a}{References}: (1)~\citet{Altamirano_2008}, (2)~\citet{1997ApJ...479L.137S}, (3)~\citet{2003HEAD....7.1738H},
(4)~\citet{Kuulkers2003}, (5)~\citet{Valenti_2007}, (6)~\citet{G_ver_2010}, (7)~\citet{G_ver_2013},
(8)~\citet{Ortolani2007}, (9)~\citet{O_zel_2009}, (10)~\citet{O_zel_2012}. \\
\tablefoottext{b}{The} normal/flat distribution of the distance to EXO 1745--248 was used for our MC sampling/Bayesian analysis. \\ 
\tablefoottext{c}{The} flat distribution of the distance to KS 1731--260 was used for both our MC sampling and Bayesian analysis. \\
\tablefoottext{d}{The} distance measurement for 4U~1608--52 and the minimum distance of 3.9 kpc are from~\citet{Guver:2008gc}. We adopt a cut off below 3.9 kpc in our MC sampling, however, a Gaussian distribution was used in~\citet{Steiner_2010}. For our Bayesian analysis, we use the posterior distribution in \citet{Ozel_2016}, i.e., 3.63 $\pm$ 0.29. \\
}
\end{table*}

Because the flux measured at the touchdown moment is still the local Eddington flux, it can be expressed as a function of mass and radius of the neutron star, the distance to the target, and the opacity of the photosphere. After the touchdown, the flux and the temperature continue to decrease and reach quasi-equilibrium while the photospheric radius does not change. 
The flux and temperature measured at this later quasi-equilibrium phase provide the effective area 
of the observed X-ray emission which is also a function of mass and radius of the neutron star and the distance to the target. 
Given two measured quantities, the touchdown flux and effective area, the mass and radius of the neutron star can be 
obtained simultaneously when the additional observation information such as the distance to the XRBs is available. 
This idea has been suggested 
to estimate the mass and radius of the neutron star in the PRE XRBs and the first  
relatively accurate measurement was made for the neutron star in EXO 0748--676\footnote{Note that in this measurement, the observed gravitational redshift was used instead of the distance.} \citep{Ozel:2006bv}. 

Note that the above method is based upon the assumption that  
XRBs occur on the entire surface of the neutron star. 
There are a couple of theoretical and observational arguments that support this assumption. 
First, with a numerical model, \cite{Spitkovsky:2001ev} proposed that the time scale of spreading 
the thermonuclear explosion on the entire surface of a rotating neutron star 
is considerably shorter ($\ll$ 1s) than that of the XRB itself, i.e., 10 to 100 seconds. 
The other argument is based upon the weak magnetic field of the neutron stars in LMXBs which is supported by the observations that do not show strong pulsations in the steady state. If the magnetic field of the neutron star were strong, it would prevent the nuclear burning which causes XRB from occurring on the entire surface of the neutron star. 
However, some uncertainties in the spectral analysis that 
contributes to the measurement of the effective area still remain. The observed blackbody spectrum often deviates from that of the true blackbody due to atmospheric effect. If the emitting area were not perfectly spherical, the measurement of the effective area could also be altered. 
In order to address these uncertainties, a parameter called "color-correction factor" is introduced in the equation for the 
effective area. 

\begin{figure*}
\centering
\includegraphics[draft=false,width=1.0\textwidth]{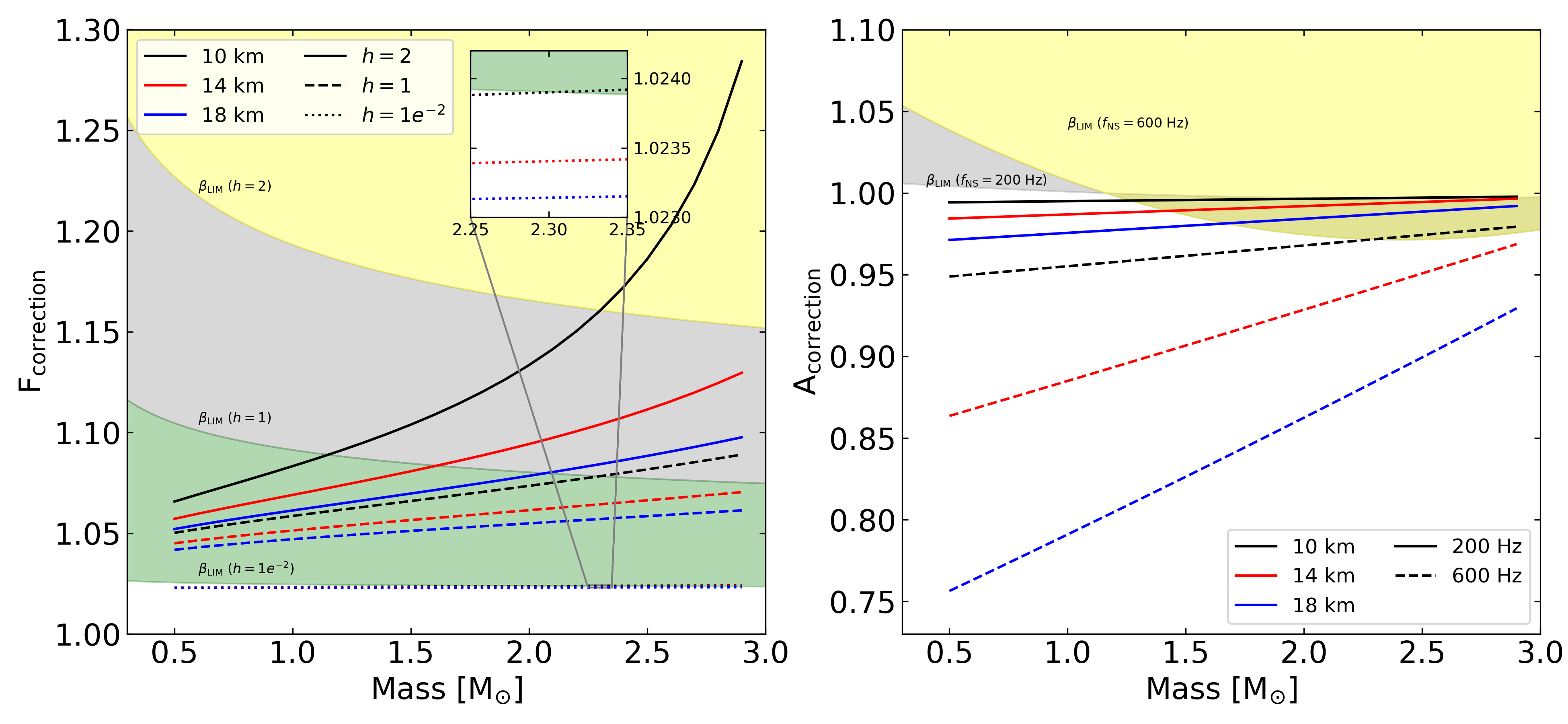}	
\caption{Correction terms, equations (\ref{Eq_A_corr}) and (\ref{Eq_F_corr}), are plotted as a function of mass for a range of radius which is indicated with different colors in both panels. For $F_{\rm corr}$ (left panel), plots are drawn with three fixed values of $h = 2R/r_{\rm ph}$.  In case of $h=0.01$, all three lines are almost overlapped (see the inset).
$A_{\rm corr}$ is not affected by $h$ but by $f_{\rm NS}$ for which two values are selected (right panel).
In both panels, regions for the causality limit ($\beta = GM/Rc^2 < 1/ 2.94$) are indicated with different colors for different values of $h$ and $f_{\rm NS}$. The regions above the boundaries are not allowed physically.
} \label{fig:correction_term}
\end{figure*}

By paying further attention to the color-correction factor or the deviation from the blackbody spectrum during the cooling phase (i.e., after the touchdown), \citet{Suleimanov:2010th} suggested a new method called "cooling tail method". In this method, the spectral evolution during the cooling phase which depends on the radiated flux (or luminosity) is compared with the atmosphere model of a neutron star which takes the surface gravity and chemical composition as input parameters. The comparison provides an alternative method to measure the Eddington flux (and effective area) although the Eddington flux measured in this method was smaller than the measured touchdown flux for 4U 1724--307 which was analyzed as a test case. More recently, the cooling tail method was developed further into "direct cooling tail method" in which the direct comparison of the observed spectral evolution with the atmosphere model provides the mass and radius of the neutron star without inferring the Eddington flux and the effective area \citep{Suleimanov:2016llr}. They applied the direct cooling tail method to SAX J1810.8--2609 which shows PRE and compared the two (standard and direct cooling tail) methods. The direct cooling tail method was evolved further with the Bayesian approaches. 
By applying the Bayesian fitting between the model spectra predicted from the direct cooling tail model and the observed spectra from the five hard-state X-ray bursts of 4U 1702--429, \citet{Nattila:2017wtj} derived the probable values for the mass and the radius of the neutron star in this LMXB, distance, and hydrogen mass fraction.    
Despite the success of the cooling tail method, however, the validity of the atmosphere model in use is limited. For example, by comparing the coherent oscillations observed during the quiescent phase to the XRB observations in 4U 1728--34, \citet{Zhang:2015nda} found that the atmosphere model could only explain the observed spectrum during the cooling phase of a confined subset of XRBs which do not show oscillations during the quiescent phase.

When a powerful PRE XRB occurs, the temperature of the radiating region is high enough 
to fully ionize both H and He. As a result, 
the opacity of the photosphere is dominated by Thomson scattering and determined by the chemical composition of the photosphere such as  
$\kappa = 0.2(1+X) \rm ~cm^2~g^{-1}$, where $X$ is the hydrogen mass fraction in the H-He plasma.
There have been both observational and theoretical efforts to constrain $X$. The observational efforts were based upon the identification of the companion stars including their stellar types which could provide the information on the chemical composition of the accreted material. However, the chemical composition of the accreted material is not necessarily identical to that of the photosphere because of the steady-state burning of hydrogen into helium that takes place during accretion. The theoretical efforts were based upon the ignition model that explains how XRBs are empowered via the nuclear runaway reactions. According to this model, the ratio of the integrated persistent flux to the burst fluence  can constrain the fuel type, i.e., hydrogen, helium, and their mixture because burning different fuel results in different energy production efficiency. Thus, comparison of the observed ratio to the prediction could constrain the composition of the fuel although the chemical composition of the fuel is neither necessarily identical  to the photospheric composition. As a result, it is quite uncertain yet to accurately constrain $X$, the hydrogen mass fraction in the photosphere, which appears in the equations to determine the mass and radius of the neutron star in a LMXB which shows PRE XRBs.

Since the first accurate measurement of \citet{Ozel:2006bv}, several LMXBs that show PRE XRBs have been analyzed for the simultaneous estimation of the neutron star mass and radius with various methods, for example, (1) statistical methods to solve the Eddington flux equation and the apparent angular area equation (including the additional effects of spin and color temperature) and (2) cooling tail method.  
A recent review is available in \citet[][also see the references therein]{Ozel:2016oaf}. 
Note that different groups used different methods depending on the assumptions they made and the observational values they used. As a result, different values of mass and radius were obtained for the neutron stars even from the identical PRE XRBs. 
However, most methods commonly dealt with two uncertain quantities, opacity and color-correction factor, which are often considered as model-dependent parameters although it is possible to constrain them observationally in principle. Because the observational constraints on these quantities have not been strong yet, previous works based upon the statistical methods such as a Monte Carlo (MC) sampling (also called a frequentist analysis) and a Bayesian analysis often adopted uniformly distributed values in the allowed range.

In this work, 
we take into account the possibility that touchdown occurs away from the neutron star surface, which was proposed by \citet{Steiner_2010}, and revisit the estimation of the mass and radius of a neutron star in LMXB that shows PRE XRBs by focusing on the effects of the chemical composition of the photosphere and the touchdown radius.  
We investigate whether (or how much) the estimated mass and radius of the neutron star change systematically depending on the the chemical composition of the photosphere and the touchdown radius. We also try to find whether there is any statistical trend for these two parameters including a correlation between them. For our investigation, we use both a MC sampling and a Bayesian analysis and apply these statistical methods to six LMXBs which were previously analyzed either with the Bayesian method \citep[e.g.,][]{Ozel_2016} or with the MC sampling \citep[e.g.,][three targets]{Steiner_2010}. 

In the next section, we describe the equations from which the mass and the radius of a neutron star in LMXB that shows PRE XRBs are estimated. In the same section, we also present the method of our MC sampling and Bayesian analysis. Our result and conclusion are presented in Section \ref{s03} and \ref{s04}, respectively.

\begin{figure}
\centering
\includegraphics[draft=fals(e,width=\columnwidth]{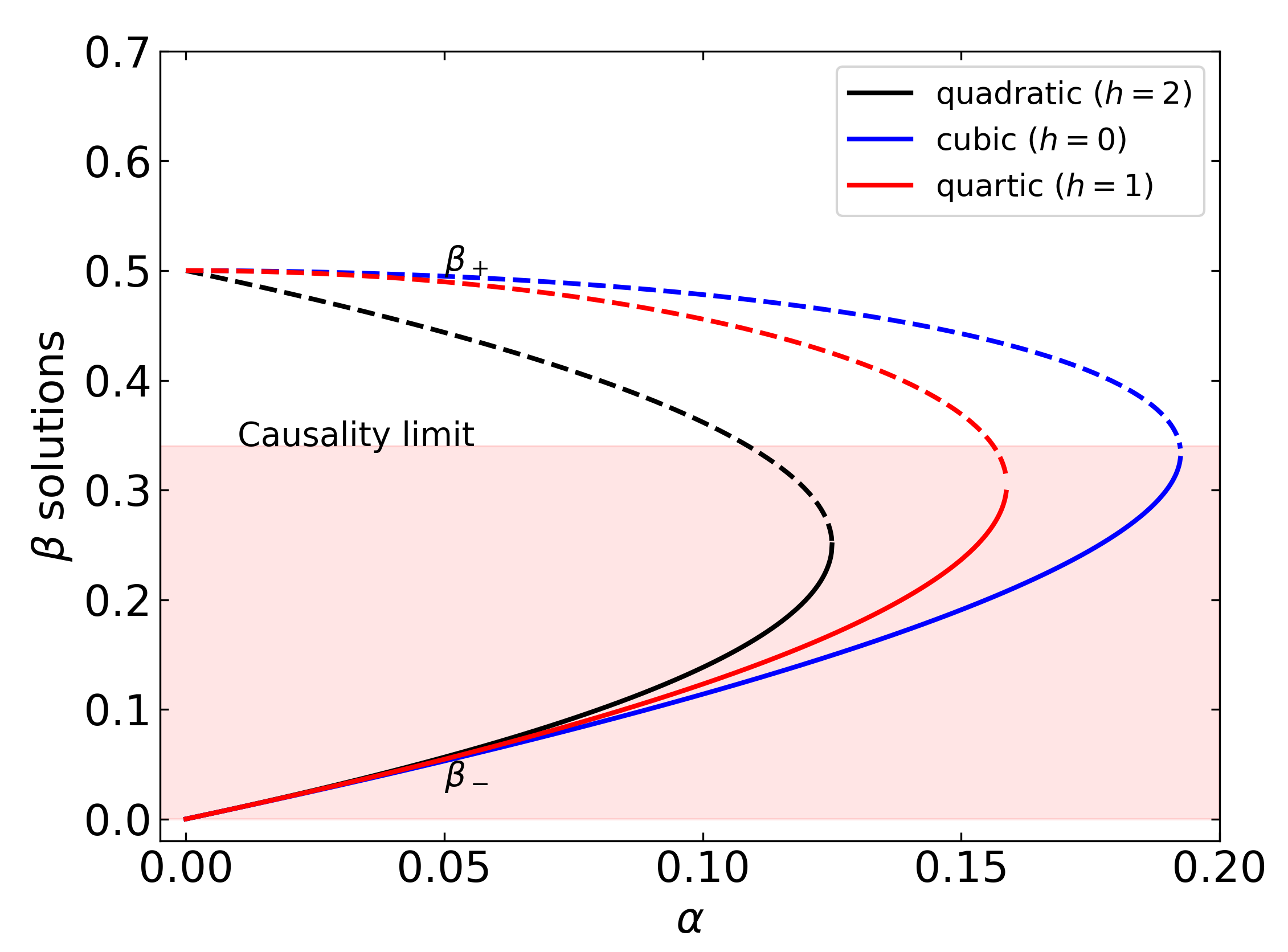}	
\caption{Physically allowed solutions ($\beta$) for equations (\ref{Eq_alp1}) to (\ref{eq:LMXB_2nd}) as a function of $\alpha$. Solid lines represent $\beta_-$ while dashed lines $\beta_+$ ( $\beta_+ > \beta_-$). 
The red shade region indicates the causality limit ($\beta < 1/ 2.94$). Solutions within this region are physically meaningful.} \label{fig:beta_solution}
\end{figure}

\begin{figure*}[t]
\begin{center}
\includegraphics[draft=false,width=0.8\textwidth]{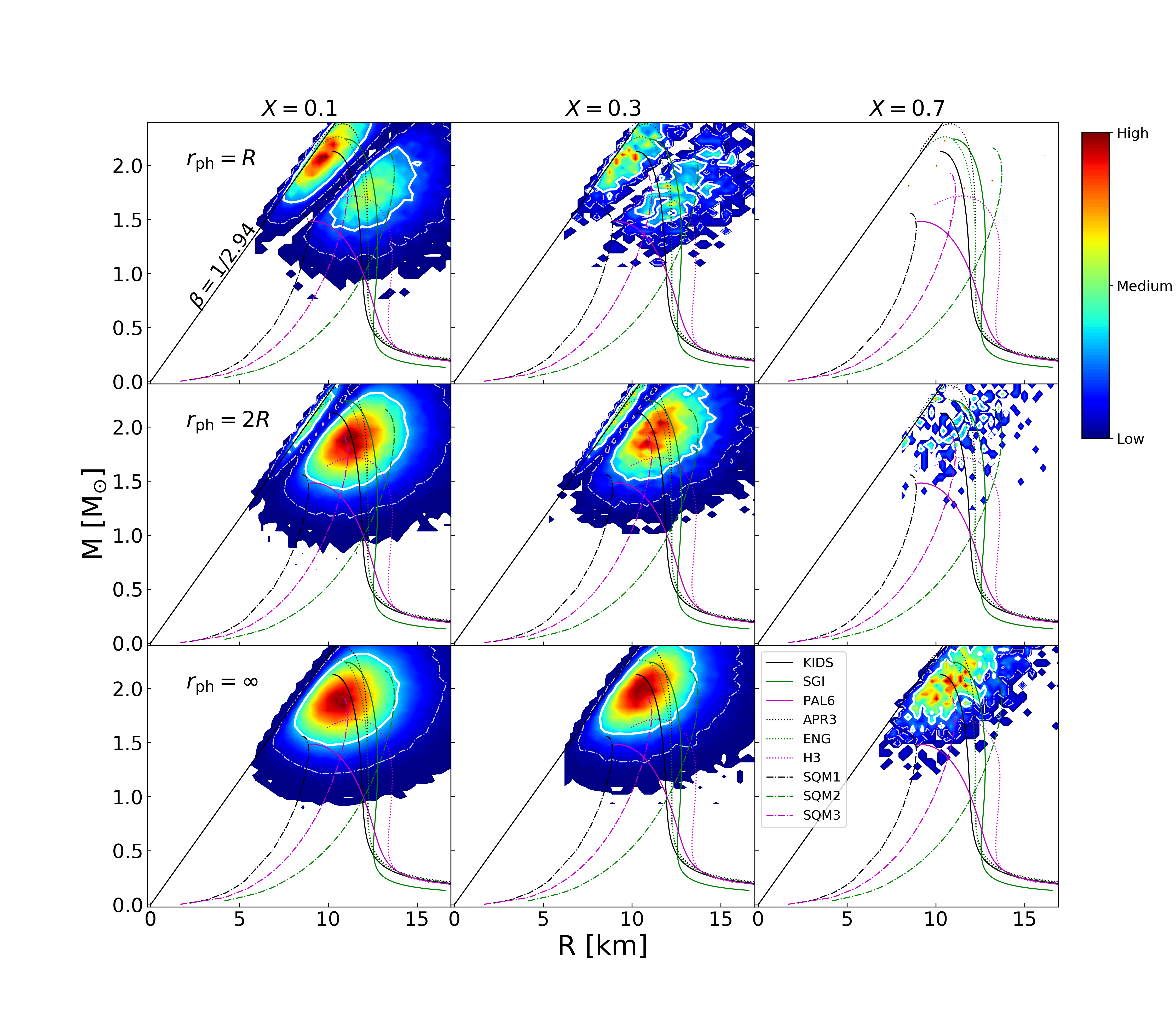}
\end{center}
\caption{Probability distributions of mass and radius obtained with our MC sampling for 4U~1820--30. Between two distance measurements for this source, we choose to use a larger one, $D=8.4 \pm 0.6$~kpc, for the result shown here. In all panels, the $68\%$ (solid white lines) and  $95\%$ (dot-dashed white lines) confidence contours are drawn. The color bar on the right reflects the relative probabilities the details of which are given in the text. 
The general relativistic causality limit ($\beta_{\rm limit} = 1/2.94$) is drawn as a black straight line.  
From top to bottom, $h=2$, $h=1$ and $h=0$, which correspond to $r_{\rm ph} = R$ (top), $r_{\rm ph} = 2R$ (middle), and $r_{\rm ph} >> R$ (bottom), respectively.  
From left to right, $X = 0.1$ (left), $X=0.3$ (middle), and $X = 0.7$ (right). 
We note that the relative probabilities $P_i$ of all distributions are normalized by $\int P_i dR dM = 1$.
For comparison, we also plot the mass--radius curves predicted from nine equation of state (EoS) models. KIDS is a recently developed EoS model which is based upon density-functional theory. KIDS incorporates a perturbation scheme for the expansion of energy density-functional \citep{Papakonstantinou:2016zpe}. Both SGI and PAL6 take a potential-method approach \citep{Li:1991px,PhysRevLett.61.2518}, but SGI is based upon the Skyrme force model more specifically. 
APR3 and ENG are a variational-method EoS and a relativistic Brueckner-Hartree-Fock EoS, respectively \citep{Akmal:1998cf,Engvik:1995gn}. All these five models include only regular nucleons, protons and neutrons. 
However, H3 considers the hyperon contribution \citep{Lackey:2005tk} while SQM1,2,3 include the effect of quarks \citep{Prakash:1995uw}. These two EoS models are based upon relativistic mean-field theory. We note that five EoS models in our selection (KIDS, SGI, APR3, ENG, and SQM2) can predict the neutron star mass larger than $2 ~\msun$ which is now a standard observational constraint \citep{Demorest:2010bx,Antoniadis:2013pzd}. 
}\label{fig:4U1820-30}
\end{figure*}

\begin{figure*}[t]
\centering
\includegraphics[draft=false,width=0.8\textwidth]{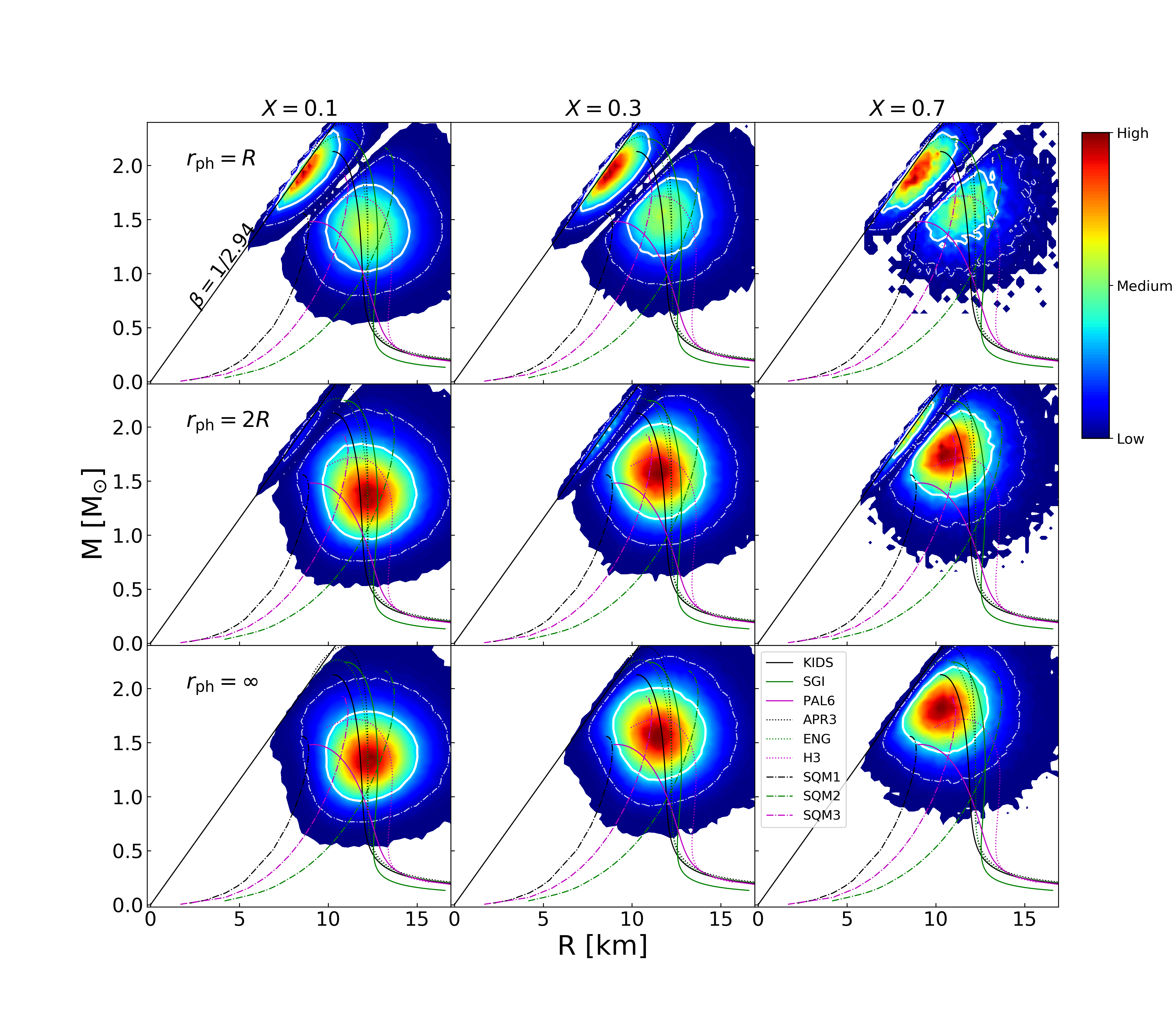}
\caption{Same as Figure \ref{fig:4U1820-30} for SAX J1748.9--2021.}
\end{figure*}

\begin{figure*}[t]
\centering
\includegraphics[draft=false,width=0.8\textwidth]{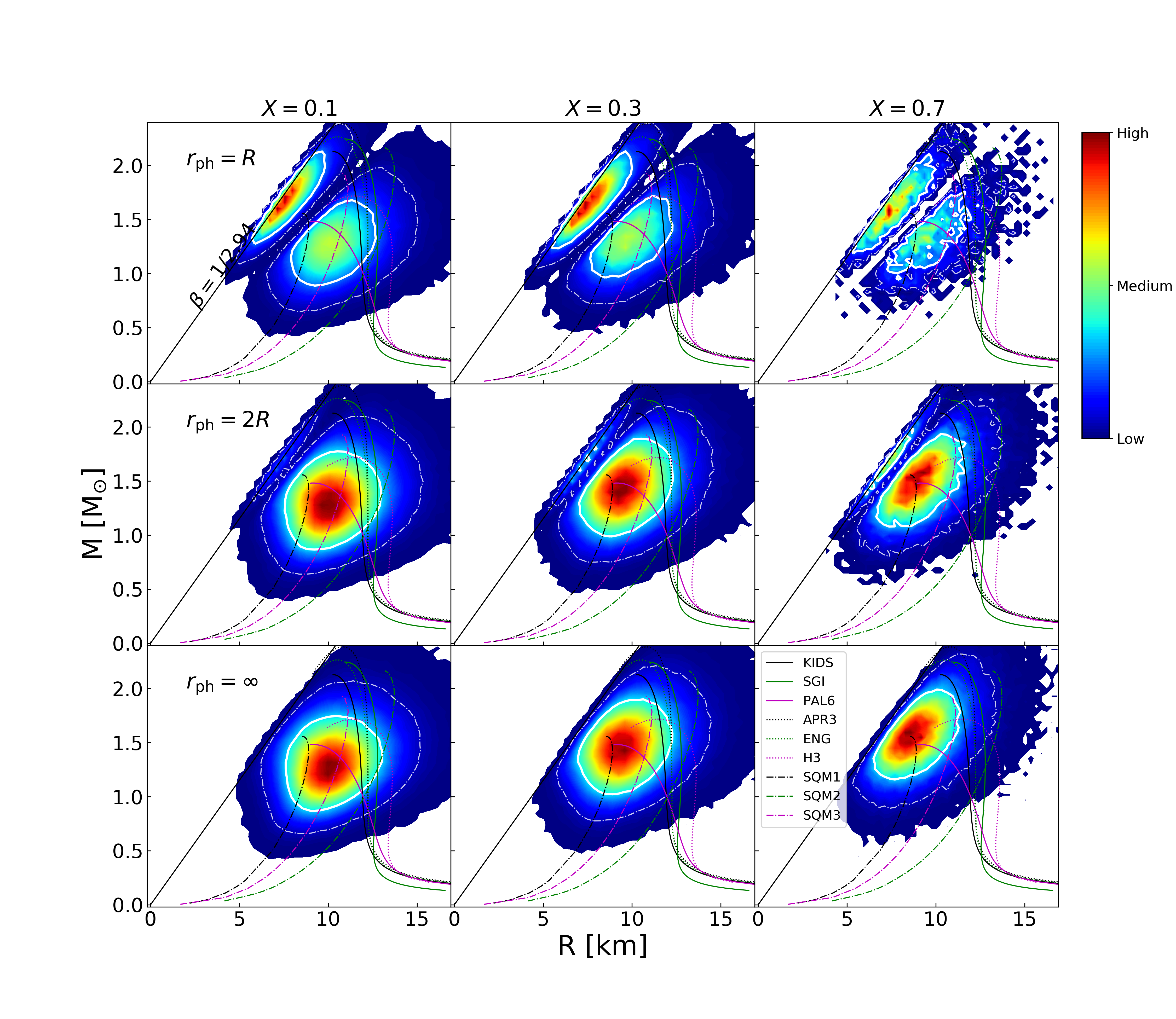}
\caption{Same as Figure \ref{fig:4U1820-30} for EXO1745--248.}
\end{figure*}

\begin{figure*}[t]
\centering
\includegraphics[draft=false,width=0.8\textwidth]{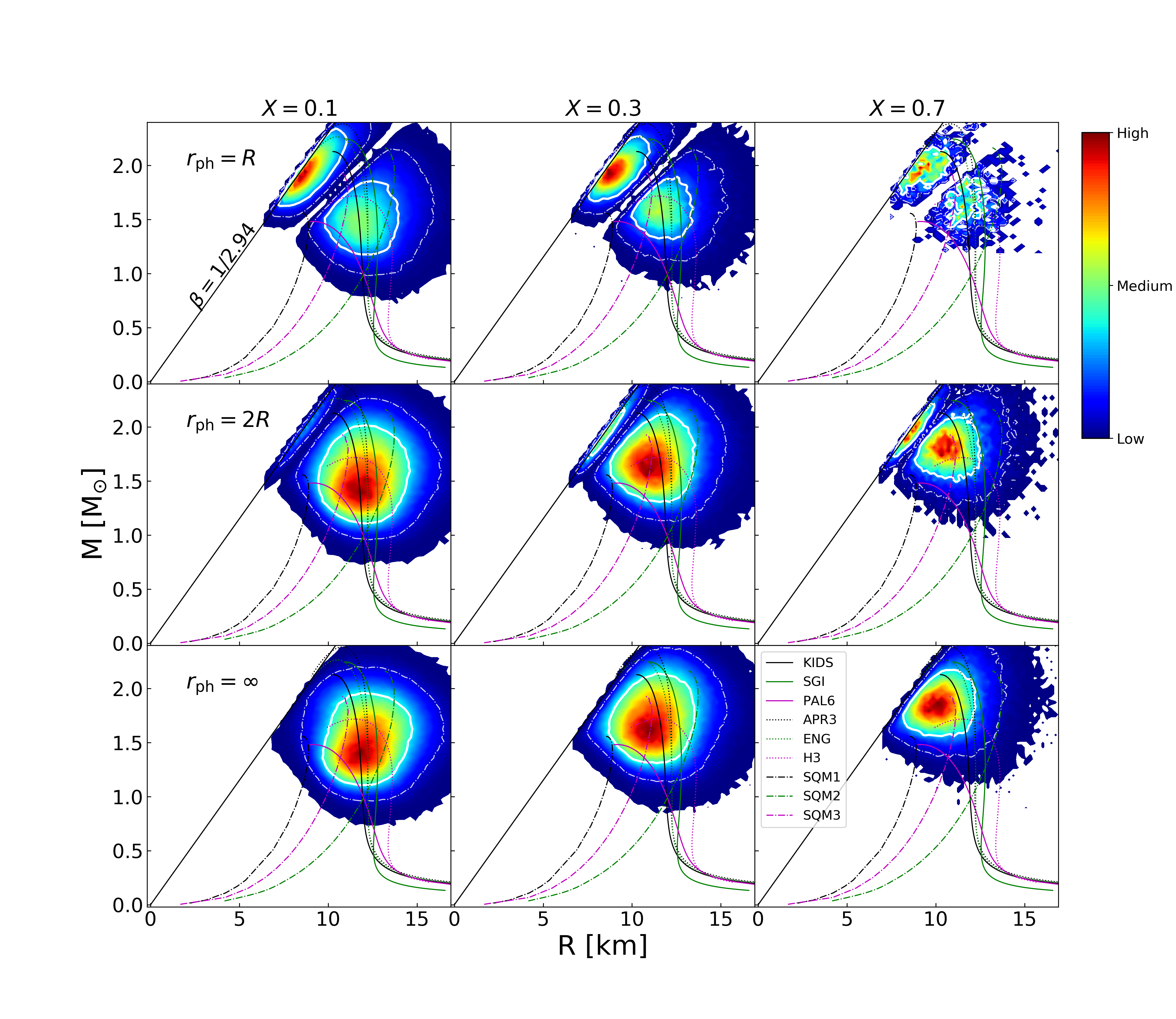}
\caption{Same as Figure \ref{fig:4U1820-30} for KS1731--260.}
\end{figure*}

\begin{figure*}[t]
\centering
\includegraphics[draft=false,width=0.8\textwidth]{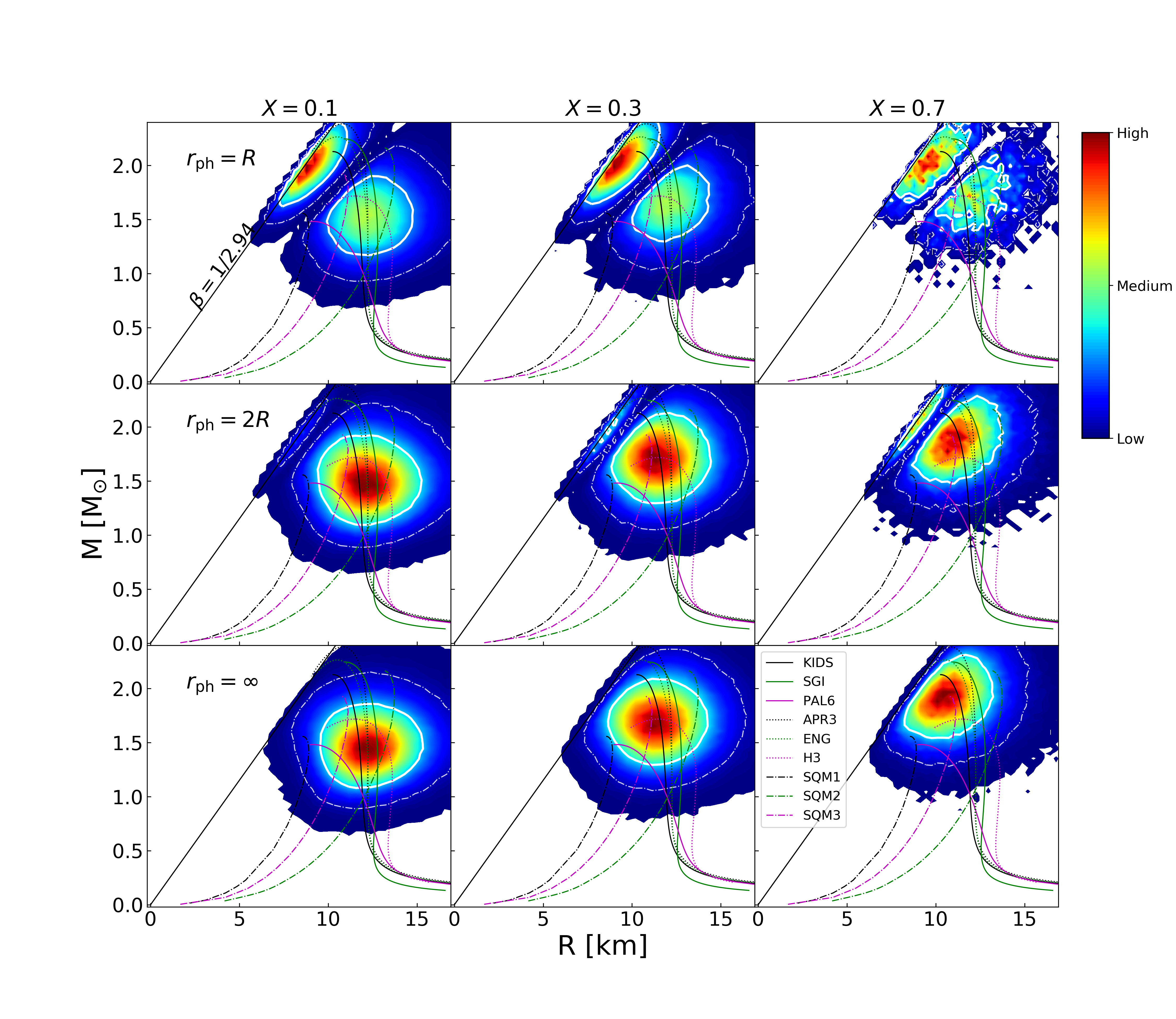}
\caption{Same as Figure \ref{fig:4U1820-30} for 4U 1724--207.}
\end{figure*}

\begin{figure*}[t]
\centering
\includegraphics[draft=false,width=0.8\textwidth]{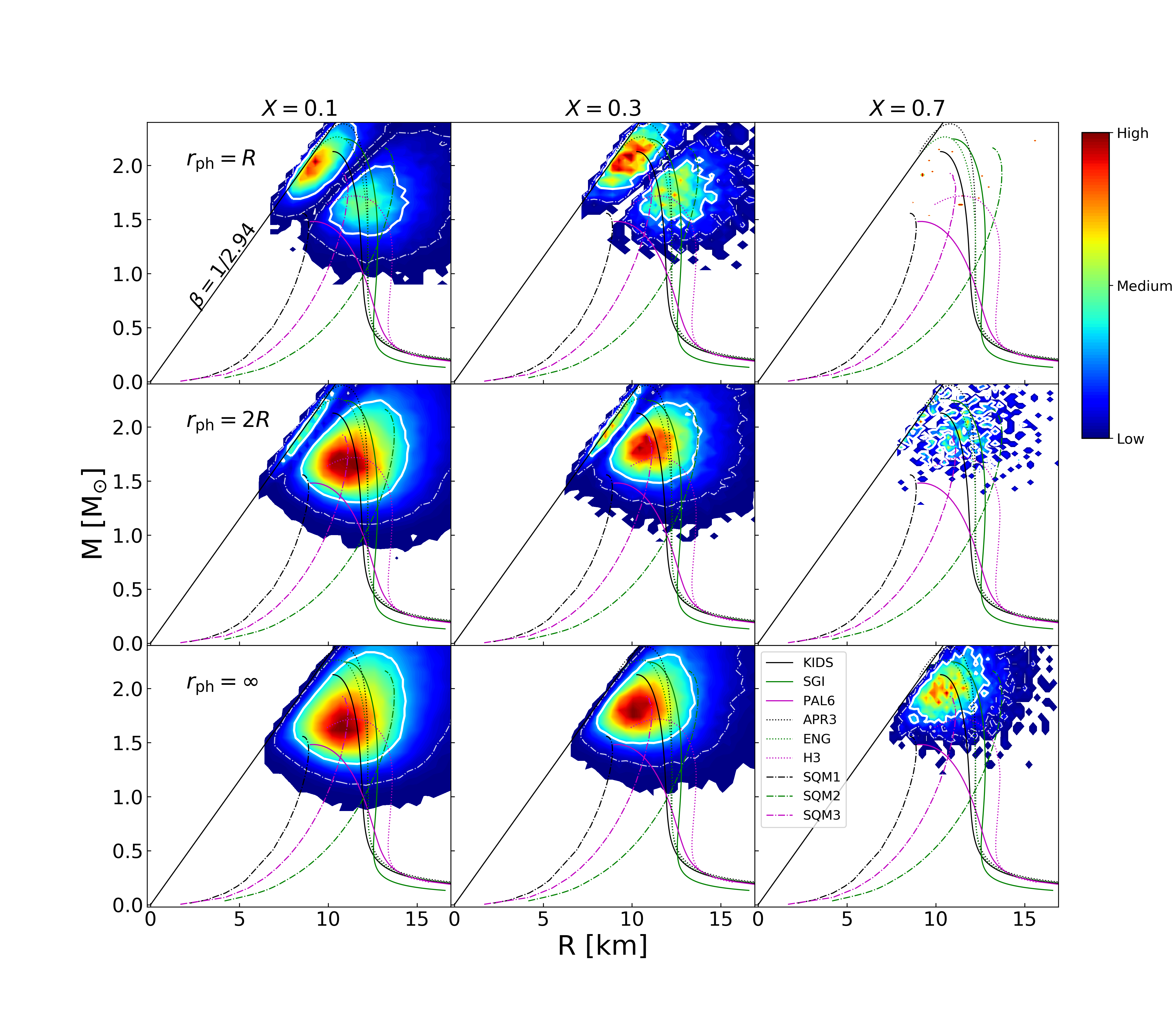}
\caption{Same as Figure \ref{fig:4U1820-30} for 4U 1608--52.}
\label{fig:4U1608-52}
\end{figure*}

\begin{figure*}[t]
\centering
\includegraphics[draft=false,width=0.9\textwidth]{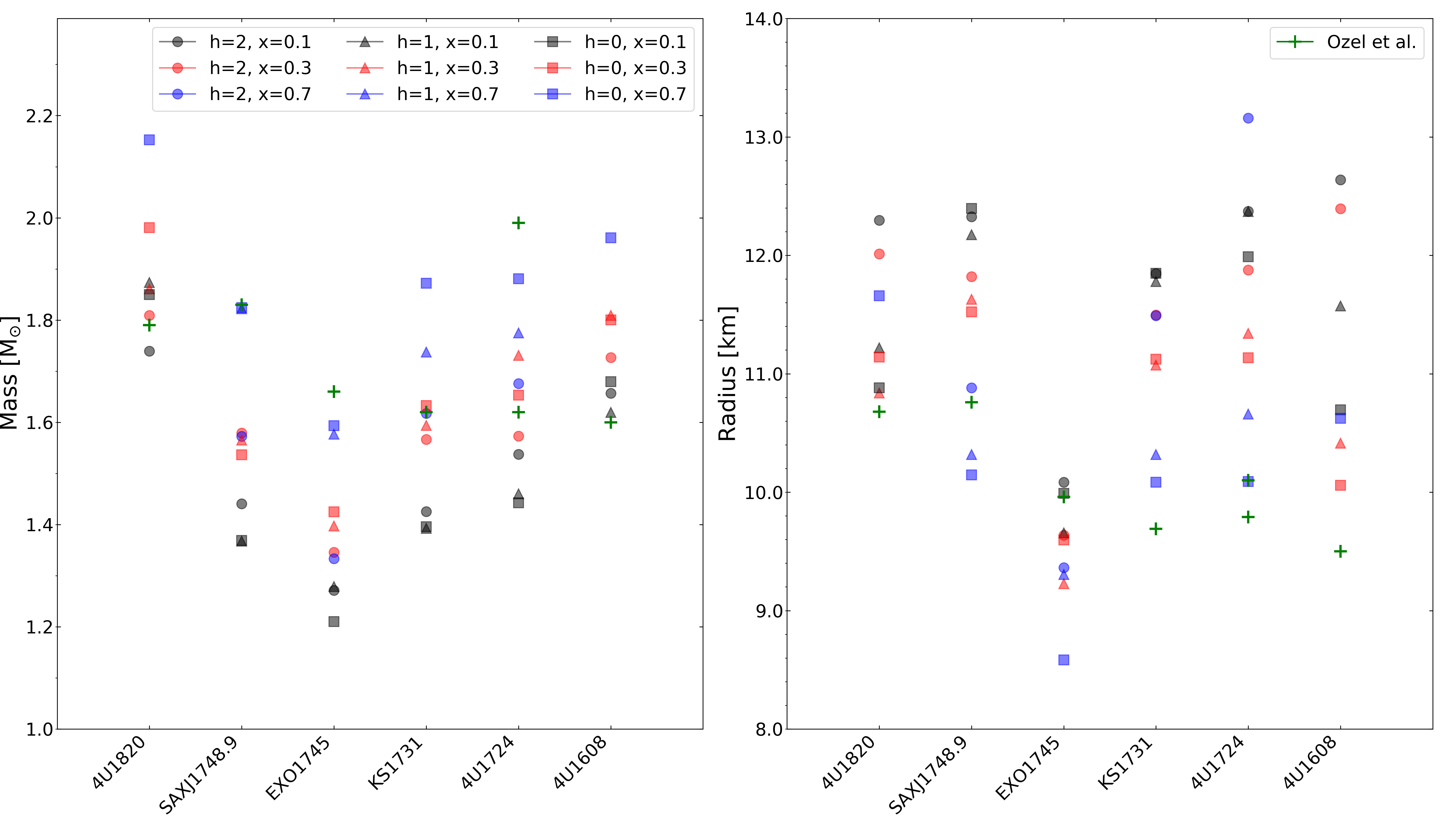} 
\caption{Most probable values of mass (left panel) and radius (right panel) for the six sources that we obtained from our MC sampling for each of the values
of $X$ and $h$. The results from \citet{Ozel_2016} are also included for comparison. As in Figure \ref{fig:4U1820-30}, we choose to use a larger distance only for 4U~1820--30.}
\label{fig:most_mr}
\end{figure*}

\section{Method}\label{s02}

\subsection{Equations}

Two quantities, touchdown flux ($\Ftdi$) and apparent angular area ($A$), are obtained from 
the observed light curve and spectrum of PRE XRBs and expressed 
as functions of mass ($M$) and radius ($R$) of the neutron star, distance to the target ($D$), 
opacity ($\kappa$), and color-correction factor ($\fc$) as 
in the following two equations, respectively. 
\be
\Ftdi =\dis \frac{GMc}{\kappa D^2}\biggl(1-\frac{2GM}{Rc^2}\biggr)^{1/2}, \label{edd}
\ee
\be
A \equiv \frac{\Finf} {\sigma \Tbbi^4 }= \fc^{-4} \frac{R^2}{D^2}\biggl(1-\frac{2GM}{Rc^2}\biggr)^{-1},\label{A}
\ee
where $c$, $G$, and $\sigma$ are speed of light, gravitational constant, and Stefan$-$Boltzmann constant, respectively. 
The $(1-\frac{2GM}{Rc^2})$ terms appear in order to include the effect of general relativity. 
$\Ftdi$ and $\Finf$ are measured from the 
light curve at the touchdown moment and at the later quasi-equilibrium phase after the touch down, respectively 
and $\Tbbi$ is from the spectrum (at the later quasi-equilibrium phase). 
The $\infty$ subscript means that the quantities are measured at the Earth.

In equation (\ref{edd}), touchdown is assumed to occur on the surface of the neutron star.
This assumption was made by \citet{Ozel:2006bv} who estimated the mass and radius of the neutron star in EXO 0748--676 relatively accurately for the first time. 
This assumption was also used for their later works, which estimated 
the masses and the radii of the neutron stars in other PRE XRBs \citep{{Guver:2008gc},{Guver:2010td},{O_zel_2009},{Ozel:2010fw}}. 
However, 
\citet{Steiner_2010} pointed out that this assumption resulted in poor acceptance rates in the MC simulations 
and suggested an alternative possibility that touchdown could occur away from the surface of the 
neutron star. In this assumption, $R$ in equation (\ref{edd}) is replaced with $\rph$, the photospheric radius, 
at which touchdown occurs, and equation (\ref{edd}) becomes 
\be
\Ftdi =\dis \frac{GMc}{\kappa D^2}\biggl(1-\frac{2GM}{\rph c^2}\biggr)^{1/2}~. \label{edd2}
\ee

Later, \citet{Ozel_2016} modified equations (\ref{edd}) and (\ref{A}) further as follows by considering temperature correction on the touchdown flux and spin frequency ($f_{\rm NS}$) on the apparent angular area ($A$). 
\be 
F_{\rm TD,\infty} &=& {GMc \over{\kappa D^2}} \left(1 - {{2GM}\over{Rc^2}}\right)^{1/2} \nonumber \\
&&  \left[1 + \left({kT_{\rm c} \over{38.8 \,{\rm keV}}}\right)^{a_g}\left(1 - {{2GM}\over{Rc^2}}\right)^{-a_g/2} \right] ~, \label{edd3}
\ee
where
\be
a_g = 1.01 + 0.067 \log \left(g_{\rm eff} \over {10^{14}\,{\rm cm}\,{\rm s}^{-2}}\right)~,
\ee
\be
g_{\rm eff} = {GM \over R^2}\left(1 - {{2GM}\over{Rc^2}}\right)^{-1/2}~,
\ee
and the color temperature ($T_{\rm c}$) at touchdown is given 
\be
T_{\rm c} = f_{\rm c}\left(g_{\rm eff} c \over {\sigma \kappa}\right)^{1/4} = f_{\rm c} \left( GMc \over {\sigma \kappa R^2} \right)^{1/4} \left( 1 - {2GM \over {Rc^2}} \right)^{-1/8}~. \label{Eq_Tc}
\ee
\be
A &=& f_{\rm c}^{-4}{R^2 \over {D^2}}\left(1 - {{2GM}\over{Rc^2}}\right)^{-1}  
\left\{1+\left[ \left(0.108 - 0.096 {M \over \msun}\right) \right. \right. \nonumber \\ 
&& \left. + \left(-0.061+0.114{M \over\msun}\right){R\over {10{\rm \,km}}} 
- 0.128\left({R\over {10{\rm \,km}}}\right)^2\right] \nonumber \\
&& \left. \left({f_{\rm NS} \over {1000{\rm \,Hz}}}\right)^2\right\}^2 ~ \label{A2}.
\ee
We note that touchdown is still occurring on the surface of the neutron star ($\rph=R$) in equation (\ref{edd3}). 

In this work, we take into account the possibility that touchdown does not occur on the surface of the neutron star and apply it to equation (\ref{edd3}). Then, equations (\ref{edd3}) to (\ref{Eq_Tc}) for the touchdown flux can be re-written as follows by replacing $R$ with $\rph$. 
\be 
F_{\rm TD,\infty} &=& {GMc \over{\kappa D^2}} \left(1 - {{2GM}\over{\rph c^2}}\right)^{1/2} \nonumber \\
&&  \left[1 + \left({kT_{\rm c} \over{38.8 \,{\rm keV}}}\right)^{a_g}\left(1 - {{2GM}\over{\rph c^2}}\right)^{-a_g/2} \right] ~, \label{edd4}
\ee
where
\be
a_g = 1.01 + 0.067 \log \left(g_{\rm eff} \over {10^{14}\,{\rm cm}\,{\rm s}^{-2}}\right)~,
\ee
\be
g_{\rm eff} = {GM \over \rph^2}\left(1 - {{2GM}\over{\rph c^2}}\right)^{-1/2}~,
\ee
\be
T_{\rm c} = f_{\rm c}\left(g_{\rm eff} c \over {\sigma \kappa}\right)^{1/4} = f_{\rm c} \left( GMc \over {\sigma \kappa \rph^2} \right)^{1/4} \left( 1 - {2GM \over {\rph c^2}} \right)^{-1/8}~.
\ee  
As in \citet{Steiner_2010}, we introduce a parameter $h$ which determines the touchdown radius as a function of $R$, $h = 2R / r_{\rm ph}$. We note that $h$ has a value between 0 and 2, which correspond to $\rph = \infty$ and $\rph = R$, respectively. Equations (\ref{A2}) and (\ref{edd4}) are solved for $M$ and $R$ when the values of touchdown flux ($\Ftdi$), apparent angular area ($A$), distance to the target ($D$), opacity ($\kappa$), color-correction factor ($\fc$), spin frequency ($f_{\rm NS}$), and $h$ are given. 

Although both $\fc$ and $X$, the two model-dependent parameters, are still uncertain, there have been both theoretical and observational efforts to constrain them. We have already mentioned the efforts for $X$ before in the introduction.   
As for $\fc$, \citet{Madej_2004} suggested values ranging from 1.33 to 1.81 
by considering the atmosphere models of a neutron star. 
This estimate is consistent with earlier calculations of \citet{{1986ApJ...306..170L},{1988ApJ...328..251E}}. 
As the flux approaches the local Eddington limit,
$\fc$ increases but rarely exceeds 1.5 for a typical neutron star.
However, in the tail of the bursts, $\fc$ becomes close to 1.33 
due to relatively low temperature \citep{Madej_2004}. 
In this work, we adopt a range of $\fc$ between 1.35/1.33 and 1.45/1.47 
with a central value $\bar{\fc} = 1.4$ and assume that $\fc$ is uniformly 
distributed within this range for our MC sampling/Bayesian analysis, which is the same as in \citet{Ozel_2016}/\citet{Steiner_2010}. We believe that choosing a slightly different range for $\fc$ does not affect the conclusion of this paper.

\subsection{Method 1: Monte Carlo Sampling}\label{M_MC}

In order to investigate the effects of the chemical composition of the photosphere (i.e., $X$) and the touchdown radius (i.e., $h$), we solve equations (\ref{A2}) and (\ref{edd4}) with fixed values of $X$ and $h$ ($X=0.1$, 0.3, and 0.7 and $h=0$, 1, and 2) while we take into account the distributions of the other variables based upon either observations or models. The exception for this is spin frequency ($f_{\rm NS}$) which is given as a fixed value if it is measured while a uniform distribution between 250~Hz and 650 Hz is used when a measured $f_{\rm NS}$ is not available. By using the MC sampling, we generate independent values of $\Ftdi$, $A$, $D$, and $\fc$ according to their distributions and solve equations (\ref{A2}) and (\ref{edd4}) for mass and radius with these generated values and fixed values of $X$ and $h$. The distributions of mass and radius are obtained as a result of the MC sampling. We can also get an acceptance rate, the ratio of the number of physically possible solutions to the total number of the MC sampling.  

In order to utilize the MC sampling method within a reasonable calculation time, it is necessary to solve equations (\ref{A2}) and (\ref{edd4}) relatively quickly. Since no analytic solution for $M$ and $R$ is found, we develop an iterative method which is faster than the full numerical methods. We note that equation (\ref{edd}) or (\ref{edd2}) is  solved for $M$ and $R$ analytically together with equation (\ref{A}).

As the first step to solve equations (\ref{A2}) and (\ref{edd4}) for $M$ and $R$, we pay attention to the fact that the additional correction terms introduced by \citet{Ozel_2016} are not large (i.e., close to $1$). Thus, equations (\ref{A2}) and (\ref{edd4}) can be re-written as
\be
A = A_{\rm orig}  ~ A_{\rm corr}~,
\ee
\be
\Ftdi = F_{\rm orig} ~  F_{\rm corr}~,
\ee
where 
\be
A_{\rm orig} = f_{\rm c}^{-4}{R^2 \over {D^2}}\left(1 - {{2GM}\over{Rc^2}}\right)^{-1} ~,
\ee
\be
A_{\rm corr} &=& \left\{1+\left[ \left(0.108 - 0.096 {M \over \msun}\right) \right. \right. \nonumber \\ 
&& \left. + \left(-0.061+0.114{M \over\msun}\right){R\over {10{\rm \,km}}} 
- 0.128\left({R\over {10{\rm \,km}}}\right)^2\right] \nonumber \\
&& \left. \left({f_{\rm NS} \over {1000{\rm \,Hz}}}\right)^2\right\}^2 ~\label{Eq_A_corr},
\ee
\be
F_{\rm orig} = {GMc \over{\kappa D^2}} \left(1 - {{2GM}\over{\rph c^2}}\right)^{1/2}~,
\ee
\be
F_{\rm corr} = 1 + \left({kT_{\rm c} \over{38.8 \,{\rm keV}}}\right)^{a_g}\left(1 - {{2GM}\over{\rph c^2}}\right)^{-a_g/2} ~.\label{Eq_F_corr}
\ee
Figure~\ref{fig:correction_term} confirms that the two correction terms are small unless a neutron star has an extremely small radius (large correction for the touchdown flux) or an extremely large radius with fast spin frequency (large correction for the apparent angular area). For a canonical value of $M = 1.4~\msun$ and $R=10$ km, the correction to the touchdown flux (with $h=2$ which corresponds to $\rph=R$) and the apparent angular area (with $f_{\rm NS}=600 ~\mbox{Hz}$) is about $10\%$ and $5\%$, respectively. 
In general, $F_{\rm corr}$ decreases as $h$ decreases, i.e., when touchdown occurs farther away from the neutron star surface although the mathematical limit of $F_{\rm corr}$ does not converge to $1$ exactly as $h \rightarrow 0$.  
However, in the physical limit, $F_{\rm corr}$ approaches $1$ when $h$ is sufficiently small because $F_{\rm corr}$ comes from modification of opacity, due to the energy (temperature) dependence of scattering cross section, which becomes negligible when $\rph$ is sufficiently large. \citet{Suleimanov:2012sq} mentioned the validity of the $F_{\rm corr}$ formula, for example, $T_{\rm c}$ in the range of 2 to 50 keV, although they did not consider the case $\rph \rightarrow \infty$ (or very large). Below $T_{\rm c} = 2$~keV, $F_{\rm corr}$ approaches $1$, i.e., modification of opacity becomes negligible. In equation (\ref{Eq_F_corr}), this can be  achieved by making $T_{\rm c} = 0$ below 2~keV. Alternatively, we can treat $T_{\rm c}$ as a parameter which measures the effect of opacity modification by allowing it to decrease smoothly instead of being $0$ abruptly below 2 keV. In this treatment, $T_{\rm c}$ can be smaller than 2~keV, depending on $\rph$, i.e., $T_{\rm c}$ can trace the effect of opacity modification as a function of the touchdown radius. We tested how small $h$ can be in order to treat $T_{\rm c}$ as a reasonable parameter and found that $F_{\rm corr}$ behaves well with $h = 10^{-2}$ to $10^{-3}$, which is good approximation for $h \approx 0$. In case of $h=0$, therefore, we use $F_{\rm TD,\infty} = {GMc}/{\kappa D^2}$ with $F_{\rm corr}=1$.

In the next step, we utilize the analytic solutions for equations (\ref{A}) and (\ref{edd2}) which are identical to $A_{\rm orig}$ and $F_{\rm orig}$, respectively. Following the procedures of \citet{Steiner_2010}, we define two new parameters $\alpha^*$ and $\gamma^*$, 
\be
\alpha^* \equiv {{\Ftdi}\over {\sqrt{A}}} {{\kappa D}\over{c^3\fc^2}} = \alpha ~ {{F_{\rm corr}(M,R,h)} \over \sqrt{A_{\rm corr}(M, R, f_{\rm NS})}}\, ~, \label{alpha_star}
\ee
\be
\gamma^* \equiv {{A c^3\fc^4}\over{\Ftdi \kappa}} 
= \gamma ~ {{A_{\rm corr}(M, R, f_{\rm NS})} \over {F_{\rm corr}(M,R,h)}} ~. \label{gamma_star}
\ee
The two parameters $\alpha$ and $\gamma$ defined and used in \citet[][their equations (16) and (17)]{Steiner_2010} are given as $\alpha = \beta \sqrt{1-2\beta}\sqrt{1-h\beta}$ and $\gamma = {R \over {\beta \left(1-2\beta\right)\sqrt{1-h\beta}}}$, respectively, where $\beta \equiv \frac{GM}{Rc^2}$.  
 The analytic solutions (i.e., $M$ and $R$) for equations (\ref{A}) and (\ref{edd2}) are found by solving the $\alpha$-equation for $\beta$ for a given constant value of $\alpha$ and then by using the $\gamma$-equation for $R$ with $\beta$ obtained from the $\alpha$-equation and a given constant value of $\gamma$. ($M$ is obtained from $\beta$ and $R$.) The $\alpha$-equation is generally a quartic equation for $\beta$ with $h$, but for the two special cases that we are considering in this work, the $\alpha$-equation becomes quadratic ($h=2$) and cubic ($h=0$) for $\beta$, respectively. Thus, we can write the $\alpha$-equations for the three cases in our consideration which correspond to different fixed values of $h$ (i.e., different touchdown radii), respectively, as follows. 
\be
&&\beta^2\left(1-2\beta\right) = {\alpha}^2\, ,   \hspace{1.8cm} {\rm for}\,\, r_{\rm ph} \gg R \,\,(h=0)\, \label{Eq_alp1} ,  \\
&&\beta^2 \left(1-3\beta + 2\beta^2\right) = {\alpha}^2 \, ,  \hspace{0.9cm}{\rm for}\,\, r_{\rm ph} = 2R \,\,(h=1)\, \label{Eq_alp2},  \\
&&\beta\left(1-2\beta\right) = \alpha\, , \hspace{2.08cm}{\rm for}\,\, r_{\rm ph} = R  \,\,(h=2)\, .\label{eq:LMXB_2nd} 
\ee
The solution behaviors for equations (\ref{Eq_alp1}) to (\ref{eq:LMXB_2nd}) such as constraints for the existence of physical solutions together with the causality limit ($\beta < 1/2.94$) were discussed in \citet{Steiner_2010} who considered these behaviors/constraints in their MC sampling. 
For example, $\alpha < 3^{-3/2} = 0.192$, $q^2 + p^3 < 0$ [with $p=\frac{1}{6} \left( \alpha^2 - \frac{1}{24} \right)$ and $q=-\frac{1}{12} \left( \frac{1}{144} - \frac{11}{32} \alpha^2 \right)$], and $\alpha < 1/8$ provide constraints for the existence of physically allowed solutions for equation (\ref{Eq_alp1}), (\ref{Eq_alp2}), and (\ref{eq:LMXB_2nd}), respectively. 
Figure~\ref{fig:beta_solution} shows two physically allowed solutions for the three $\alpha$-equations, equations (\ref{Eq_alp1}) to (\ref{eq:LMXB_2nd}). We note that $\beta_\pm$ shown in Figure~\ref{fig:beta_solution} are obtained with the constraints on $\alpha$ applied.

In our iterative method, we use the formal solutions of the $\alpha$-equations, equations (\ref{Eq_alp1}) to (\ref{eq:LMXB_2nd}), the forms of which are known analytically as in \citet{Steiner_2010}. Since $\alpha(M,R) = \alpha^* \,  {A_{\rm corr}}^{1/2} /{F_{\rm corr}}$, the solution for equation (\ref{alpha_star}) can be written as 
\be
\beta(M,R) = f\left(\alpha(M, R)\right)~, \label{beta_star} 
\ee  
where $f$ is the formal solution to equation (\ref{Eq_alp1}) to (\ref{eq:LMXB_2nd}), i.e., it corresponds to one of two $\beta$ solutions for each $\alpha$-equation. Similarly, with $\gamma(M,R) = \gamma^* \, {F_{\rm corr}/A_{\rm corr}}$, 
\be
R = \alpha(M,R) \gamma(M,R) \sqrt{1-2\beta(M,R)}~. \label{Eq_R}
\ee  
From the previous step $(M_n,R_n)$, $\beta_n = {G M_n}/{R_n c^2}$ is calculated but it is not the same as $f(\alpha(M_n,R_n))$ unless $(M_n,R_n)$ is the solution. So, taking $\Delta \beta = f(\alpha(M_n,R_n)) - \beta_n$, we update $\beta$ as $\beta_{n+1} = f(\alpha(M_n,R_n))$. Similarly, if $R_n$ is not the same as 
$R=\alpha(M_n,R_n) \gamma(M_n,R_n) \sqrt{1-2\beta_n}$, we update $R$ as $R_{n+1}=\alpha(M_n,R_n) \gamma(M_n,R_n) \sqrt{1-2\beta_{n+1}}$. The update of $M$ is done as $M_{n+1} = \beta_{n+1} R_{n+1} c^2 / G$.  We repeat the same process with $(M_{n+1}, R_{n+1})$ until the following convergence condition is satisfied. 
\be
&&\left(M_{n+1} - M_n\right) /M_{n+1} < \varepsilon\, ~, \\
&&\left(R_{n+1} - R_n\right) /R_{n+1} < \varepsilon\, ~,
\ee
where we choose $\varepsilon = 10^{-6}$. Due to the fact that the effects of the correction terms are not large, which we found in the first step, the convergence of this iterative method is fast when there is a solution. We check out that our method solves equations (\ref{alpha_star}) and (\ref{gamma_star}) for ($M$, $R$) correctly by plugging ($M$, $R$) back to the equations. 

Since an iterative method is useful only for finding a solution that is known to exist within a range, it is necessary to address the existence of the solution separately. The existence of the solution is related to the condition for physically allowed solutions. 
Since the effects of the correction terms are not large, the solution behaviors for equations (\ref{alpha_star}) and (\ref{gamma_star}) are not likely to deviate much from those for the $\alpha$-equations, equations (\ref{Eq_alp1}) to (\ref{eq:LMXB_2nd}) (also see Figure \ref{fig:beta_solution}). Thus, it is reasonable to apply the same $\alpha$-criteria of equations (\ref{Eq_alp1}) to (\ref{eq:LMXB_2nd}) in order to constrain the existence of physically allowed solutions for equations (\ref{alpha_star}) and (\ref{gamma_star}).  
In our iterative method, 
we set up a reasonably wide range of mass and radius, $M = 1.5 \pm 1.0 ~\msun$ and $R=15\pm 5$ km, within which the initial guess for $(M, R)$ is randomly selected. 
If the randomly selected $(M, R)$ satisfies the $\alpha$-criterion, we use this value of $(M, R)$ as the initial guess of the iteration method. Otherwise, we move onto the next set of $(M, R)$ which is also randomly generated within the same range of $M$ and $R$. We conclude that  there is no physically allowed solution when we do not find any $\alpha(M,R)$ that satisfies the $\alpha$-criterion even if we search in a significantly large range of $M$ and $R$ with a significantly large (random) sampling number. We test and confirm the validity of our method to "identify unphysical solutions" by trying a wider range of mass and radius as well as by increasing the number of random sampling ($n_{rs}$) for the case where there is no solution. For the results presented in this work, we use $n_{rs}=1000$ within the aforementioned range of $M$ and $R$. 
We also confirm that the initial guess of $(M, R)$ which satisfies the $\alpha$-criterion always results in physically allowed solutions, which implies (or confirms self-consistently) that the effects of the correction terms are not large. We note that we apply the causality limit to the random sampling of $(M, R)$, i.e., the range of $(M, R)$ is also limited with $\beta < 1/2.94$.  
For the entire MC simulations, we carry out $10^6$ realizations for each case (with three fixed values of $X$ and $h$, respectively, for each source) to produce $\alpha^*$ and $\gamma^*$. For each case with the fixed value of $X$ and $h$, the sampling is taken from the distributions of $\Ftdi$, $A$, $D$, $f_{\rm NS}$, and $f_{\rm c}$. However, in the results which show the statistics for probable $M$ and $R$, we exclude the events that do not have physically meaningful solutions from the statistical counting.

\begin{figure*}[t]
\begin{center}
\includegraphics[width=14cm]{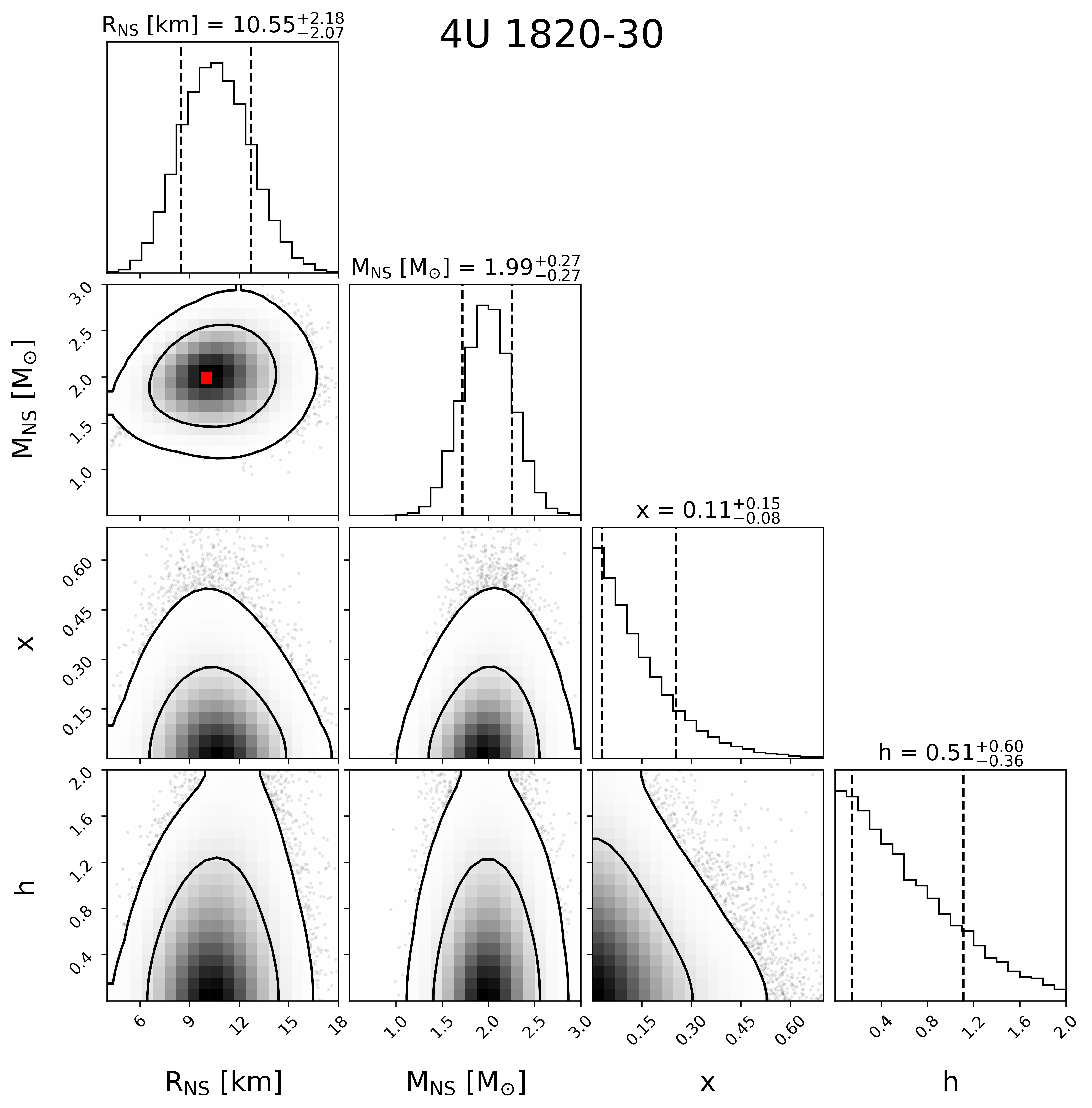}
\end{center}
\caption{Posterior distributions of neuron star mass ($M_{\rm NS}$) and radius ($R_{\rm NS}$), hydrogen mass fraction in the photosphere ($X$), and $h=2R/r_{\rm ph}$ (which determines the touchdown radius) obtained from our Bayesian analysis for 4U 1820-30. In the panels which present the posterior distributions between two model parameters, the inner and outer solid line correspond to the $68\%$ and $95\%$ confidence contour, respectively. The most probable values of mass and radius are marked with red dots in the distribution plots of mass and radius.  
In the panels which present the posterior distribution of a single model parameter as a histogram, the region between two vertical dashed lines corresponds to the $68\%$ confidence region. The numbers on top of the histogram panels present the corresponding $68\%$ confidence range around the median value. 
Since there is no spin measurement available for 4U 1820-30, we use the prior distribution of $f_{\rm NS}$ as a uniform distribution between 250~Hz and 650~Hz which is the same as in \citet{Ozel_2016}.} \label{fig-corner-1}
\end{figure*}

\begin{figure*}[t]
\begin{center}
\includegraphics[width=14cm]{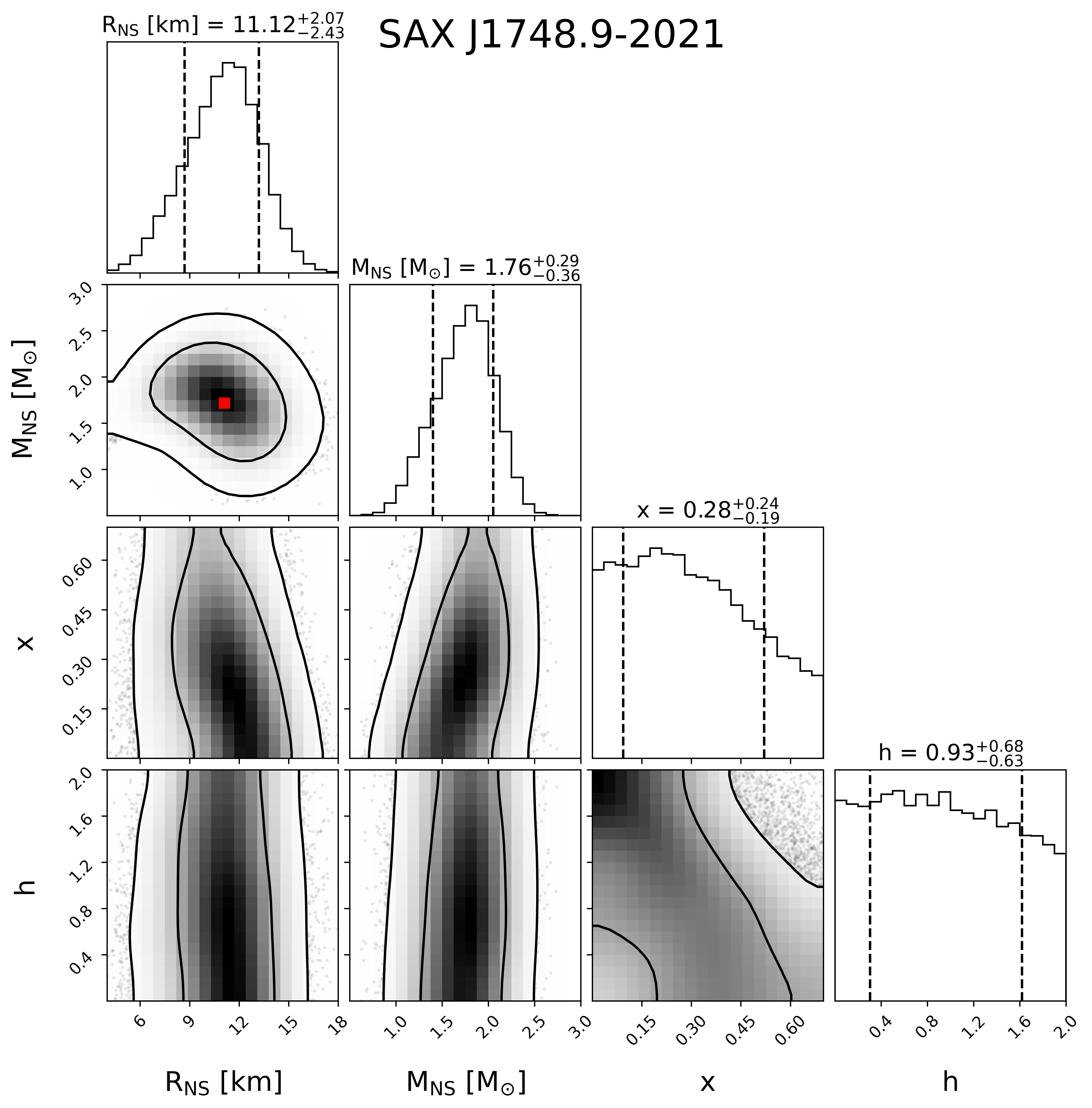}
\end{center}
\caption{Same as Figure \ref{fig-corner-1} for SAX J1748.9-2021. For this source, $f_{\rm NS}$ is available. So it was fixed at the measured value in Table \ref{ta01}.} \label{fig-corner-2}
\end{figure*}

\begin{figure*}[t]
\begin{center}
\includegraphics[width=14cm]{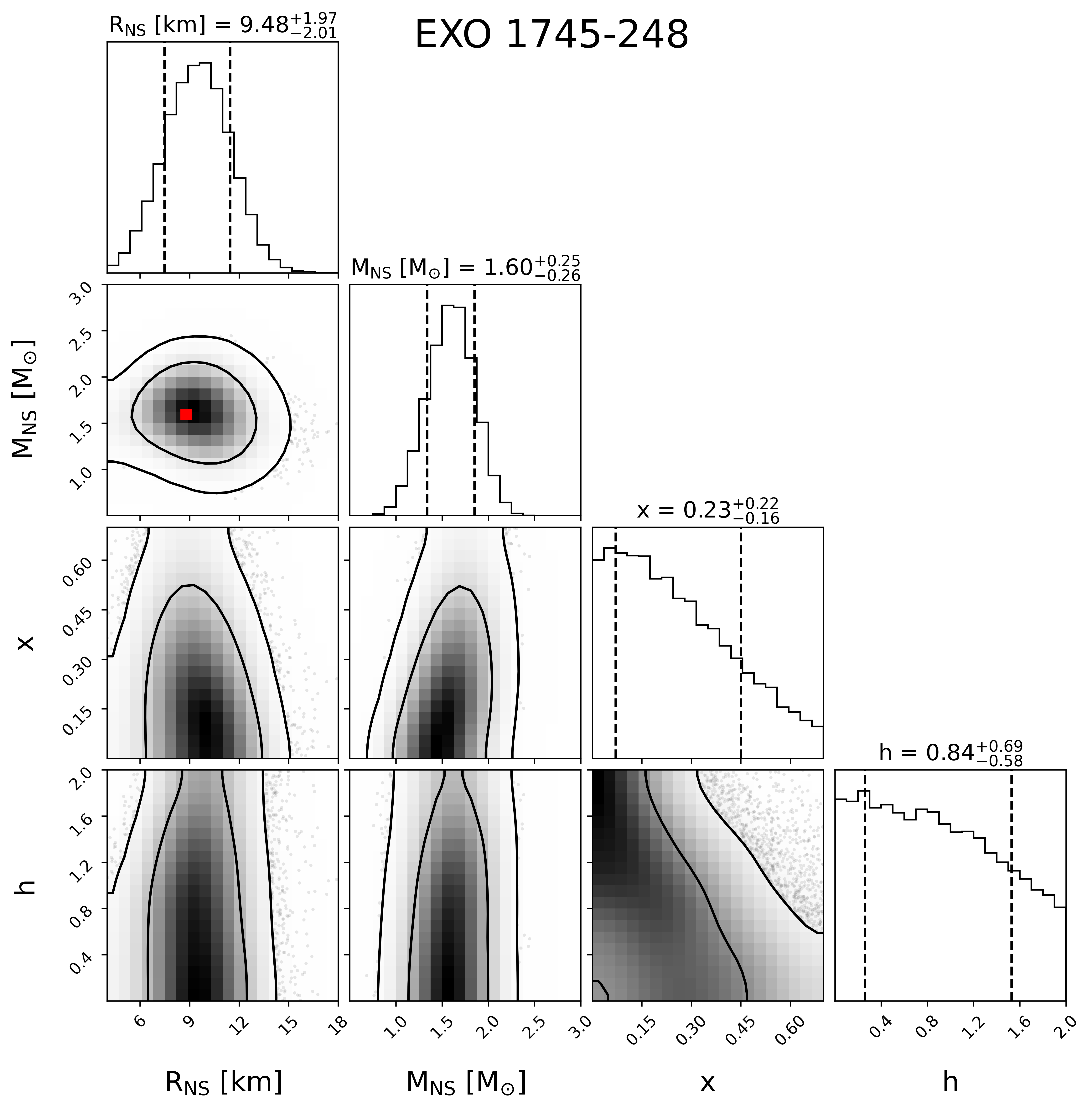}
\end{center}
\caption{Same as Figure \ref{fig-corner-1} for EXO 1745-248. Since there is no spin measurement available for this source, we use the same prior distribution of $f_{\rm NS}$ as in 4U 1820-30, i.e., a uniform distribution between 250~Hz and 650~Hz.} \label{fig-corner-3}
\end{figure*}

\begin{figure*}[t]
\begin{center}
\includegraphics[width=14cm]{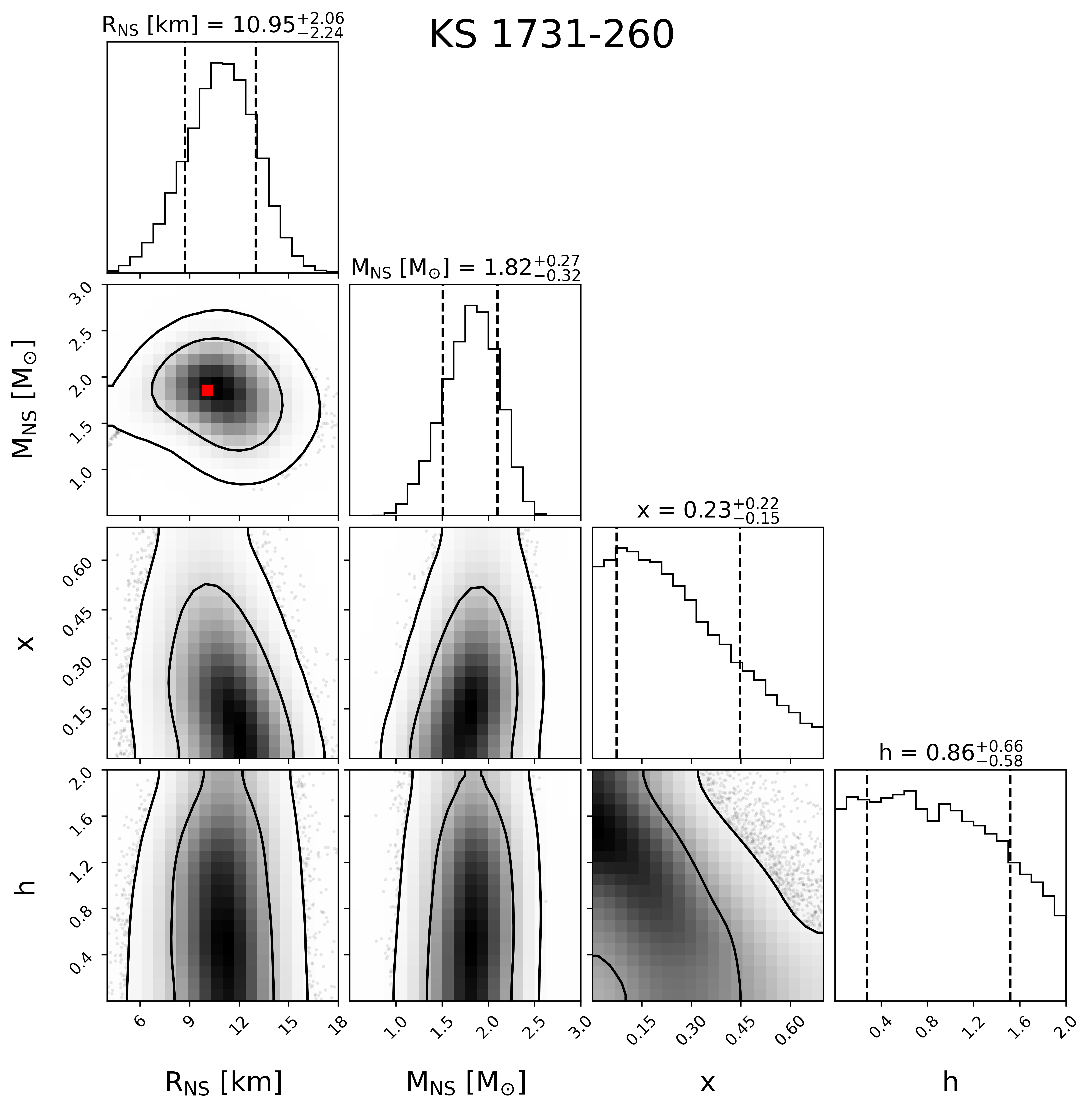}
\end{center}
\caption{Same as Figure \ref{fig-corner-1} for KS 1731-260. For this source, $f_{\rm NS}$ is available. So it was fixed at the measured value in Table \ref{ta01}.
} \label{fig-corner-4}
\end{figure*}

\begin{figure*}[t]
\begin{center}
\includegraphics[width=14cm]{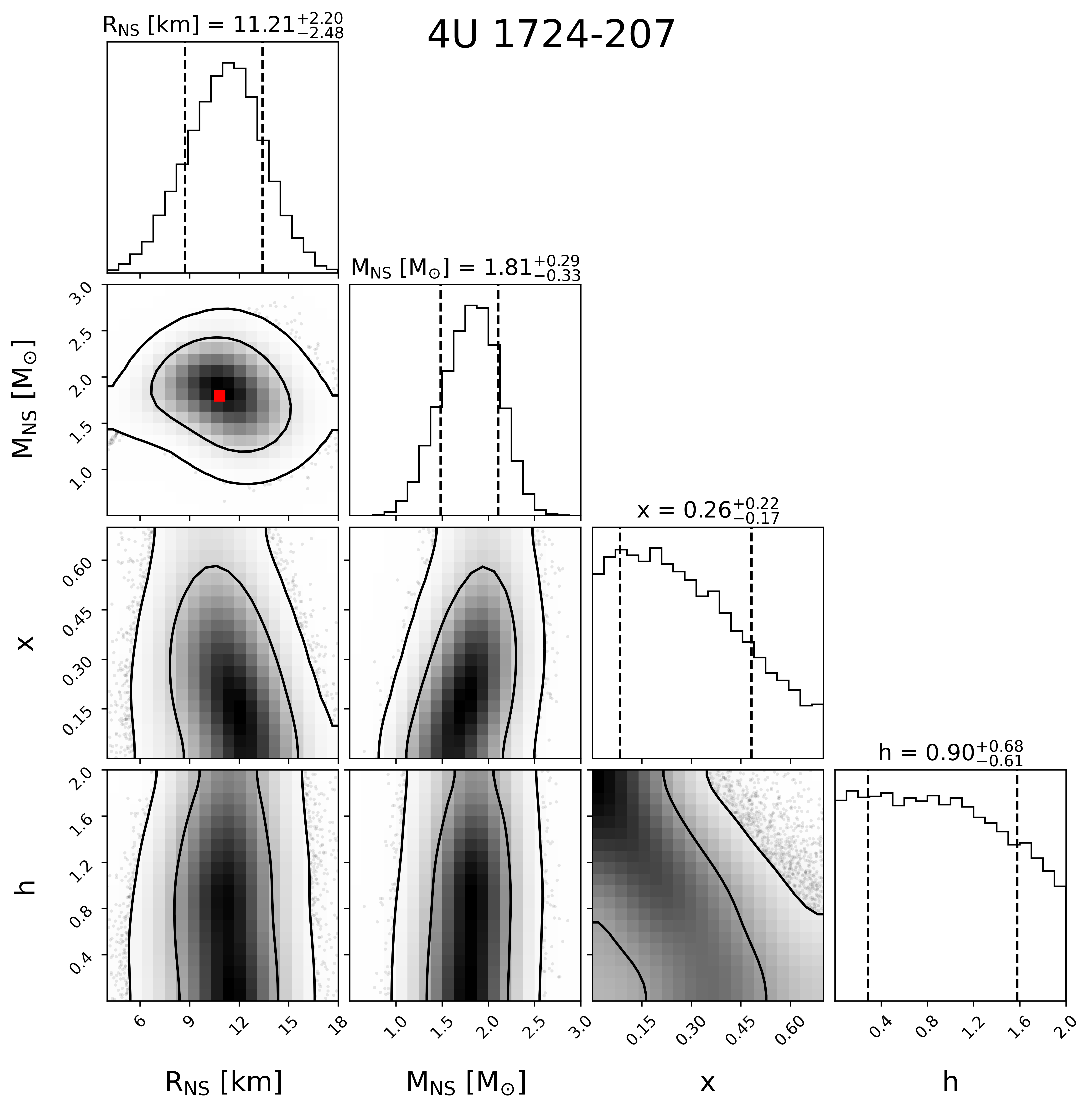}
\end{center}
\caption{
Same as Figure \ref{fig-corner-1} for 4U 1724-207. Since there is no spin measurement available for this source, we use the same prior distribution of $f_{\rm NS}$ as in 4U 1820-30, i.e., a uniform distribution between 250~Hz and 650~Hz.
} \label{fig-corner-5}
\end{figure*}

\begin{figure*}[t]
\begin{center}
\includegraphics[width=14cm]{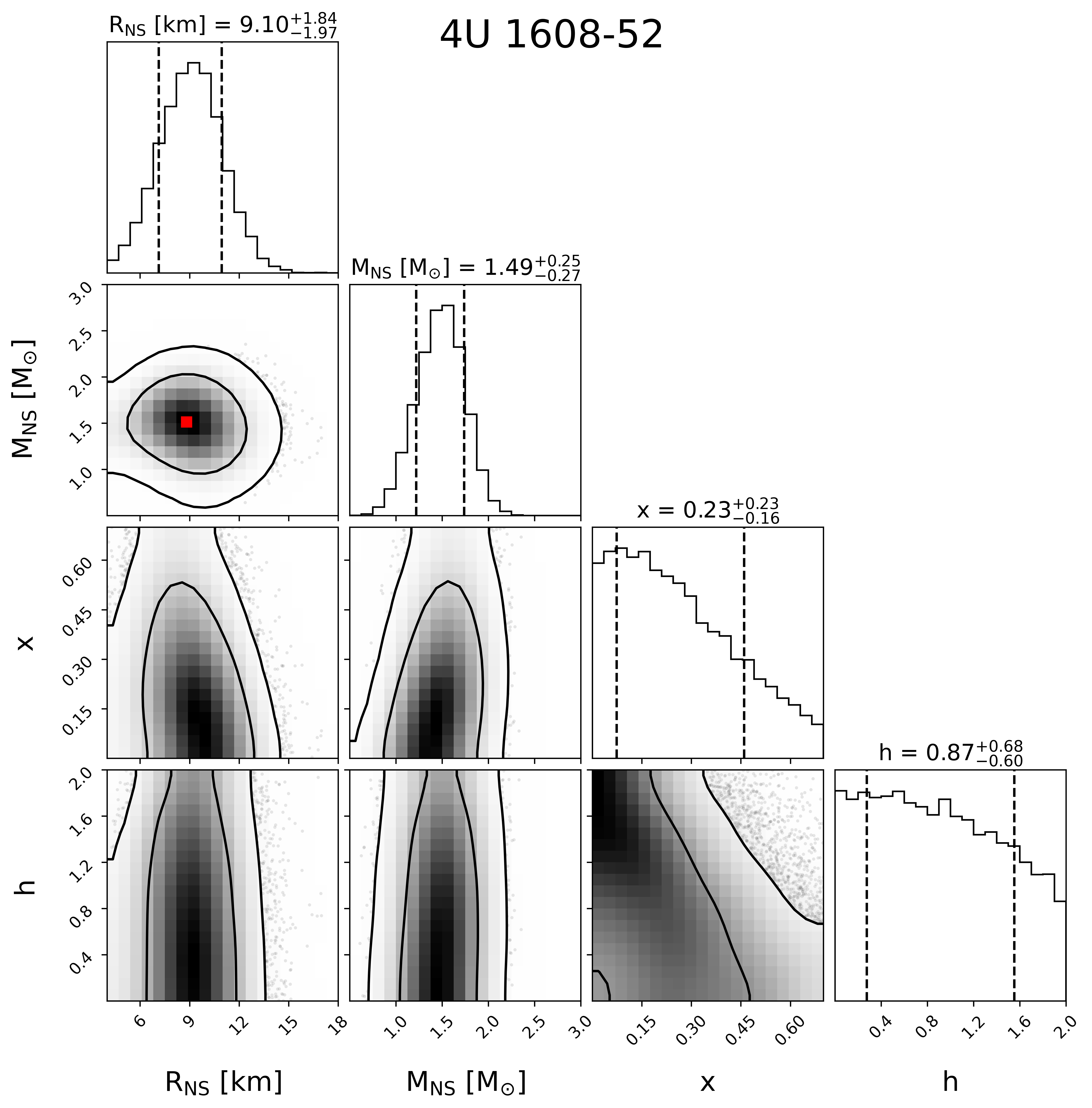}
\end{center}
\caption{Same as Figure \ref{fig-corner-1} for 4U 1608-52. 
For this source, $f_{\rm NS}$ is available. So it was fixed at the measured value in Table \ref{ta01}.
} \label{fig-corner-6}
\end{figure*}


\begin{table*}
\centering
\caption{Statistics of our MC sampling for six sources with $r_{\rm ph} = R \, (h=2)$. 
The numbers without (within) parentheses correspond to the $\beta_-$ ($\beta_+$) solution. For 4U~1820-30, we use the larger distance only, i.e., $D= 8.4 \pm 0.6$~kpc. 
 \label{table:statistics1}}
\begin{tabular}{rcccccc}
\hline \hline
\multirow{2}{*}{$r_{\rm ph} = R$}&{4U 1820--30}  & {SAX J1748.9--2021}  & {EXO 1745--248}  & {KS 1731--260}  & {4U 1724--207}  & {4U 1608--52} \bigstrut\\
&\multicolumn{6}{c}{$X=0.1$} \bigstrut\\
\hline
Sol. 		&1.2 (1.0) 			& 53.3 (21.2) 	& 30.6 (15.5) 	& 31.3 (18.4) 	& 41.2 (21.2)  	& 2.5 (2.0) \bigstrut\\
Causality 	& 0 (0.3) 			&  0 (33.9) 	& 0 (17.0) 		& 0 (14.5) 		& 0 (21.9) 		& 0 (0.7) \bigstrut\\
Unphys.	 & $\ge 98.7$  	& 46.7 (44.9) 	& 69.4 (67.5) 	& 68.7 (67.1) 	& 58.8 (56.9) 	& 97.5 (97.3)\bigstrut\\
\hline
&\multicolumn{6}{c}{$X=0.3$} \bigstrut\\
\hline
Sol. 		& $\le 0.08$ 		& 19.2 (11.3) 	& 8.0 (5.4) 	& 5.9 (4.7) 	& 10.2 (7.5)  	& 0.22 (0.19) \bigstrut\\
Causality 	& 0 (0.01)			& 0 (8.9) 		& 0 (3.3) 		& 0 (1.7) 		& 0 (3.6) 		& 0 (0.04) \bigstrut\\
Unphys. 	& $\ge 99.9$  	& 80.8 (79.8) 	& 92.0 (91.3)  	& 94.1 (93.6) 	& 89.8 (88.9)  	&$\ge  99.7$ \bigstrut\\
\hline
&\multicolumn{6}{c}{$X=0.7$} \bigstrut\\
\hline
Sol. 		& 0.0 				& 1.2 (0.9) 	& $\le 0.3$ 	& $\le 0.1$ 	& $\le 0.2$ 	& $\le 0.01$ \bigstrut\\
Causality 	& 0.0 				&0  (0.4) 		& $\le 0.1$ 	& $\le 0.01$ 	& $\le 0.05$  	& 0  \bigstrut\\
Unphys. 	& 100.0 				& 98.8 (98.7) 	& $\ge 99.6$ 	& $\ge 99.9$ 	& $\ge 99.75$ 	& $\ge 99.99$ \bigstrut\\
\hline	
\end{tabular}
\end{table*}

\begin{table*}
\centering
\caption{Same as Table~\ref{table:statistics1} for $r_{\rm ph} =2 R\, (h=1)$. \label{table:statistics2}}
\begin{tabular}{rcccccc}
\hline \hline
\multirow{2}{*}{$r_{\rm ph} = 2R$}&{4U 1820--30}  & {SAX J1748.9--2021}  & {EXO 1745--248}  & {KS 1731--260}  & {4U 1724--207}  & {4U 1608--52} \bigstrut\\
&\multicolumn{6}{c}{$X=0.1$} \bigstrut\\
\hline
Sol.		& 12.2 (1.1) 		& 90.5 (1.0) 	& 72.9 (1.7) 	& 80.3 (2.4) 	& 86.0 (1.4) 	& 16.2 (1.5)\bigstrut\\
Causality 	& 0 (11.1)			& 0 (89.6) 		& 0 (71.5) 		& 0 (78.2) 		& 0 (84.5)		&0 (14.8) \bigstrut\\
Unphys. 	&  87.8 (87.8)	& 9.5 (9.4) 	& 27.1 (26.8) 	& 19.7 (19.4) 	& 14.0 (14.1) 	& 83.8 (82.7) \bigstrut\\
\hline
&\multicolumn{6}{c}{$X=0.3$} \bigstrut\\
\hline
Sol.		& 1.8 (0.2) 		& 61.1 (2.3) 	& 37.3 (2.0) 	& 39.3 (3.1) 	& 49.5 (2.5) 	& 3.5 (0.5)\bigstrut\\
Causality	 & 0 (1.6) 			& 0 (59.0) 		& 0 (35.5) 		& 0 (36.5) 		& 0 (47.1) 		& 0 (3.0)\bigstrut\\
Unphys. 	& 98.2 (98.2) 	& 38.9 (38.7) 	& 62.7 (62.5) 	& 60.7 (60.4) 	& 50.5 (50.4) 	& 96.5 (96.5) \bigstrut\\
\hline
&\multicolumn{6}{c}{$X=0.7$} \bigstrut\\
\hline
Sol.		& $\le 0.05$ 		& 10.9 (1.0) 	& 4.0 (0.4) 	& 2.4 (0.4) 	& 4.6 (0.6) 	&$\le 0.1$ \bigstrut\\
Causality 	& $\le 0.05$ 		& 0 (10.0) 		& 0 (3.6) 		& 0 (2.0) 		& 0 (4.1) 		& $\le 0.1$  \bigstrut\\
Unphys. 	& $\ge 99.9$ 		& 89.1 (89.0) 	& 96.0 (96.0) 	& 97.6 (97.6) 	& 95.4 (95.3) 	& $\ge 99.9$\bigstrut\\
\hline
\end{tabular}
\end{table*}

\begin{table*}
\centering
\caption{Same as Table~\ref{table:statistics1} for $r_{\rm ph} \gg R \, (h=0)$. \label{table:statistics3}}
\begin{tabular}{rcccccc}
\hline \hline
\multirow{2}{*}{$r_{\rm ph} = \infty$}&{4U 1820--30}  & {SAX J1748.9--2021}  & {EXO 1745--248}  & {KS 1731--260}  & {4U 1724--207}  & {4U 1608--52} \bigstrut\\
&\multicolumn{6}{c}{$X=0.1$} \bigstrut\\
\hline
Sol.		& 31.4 (0) 		& 98.2 (0) 		& 90.1 (0) 		& 95.5 (0) 		& 97.1 (0) 		& 31.5 (0)\bigstrut\\
Causality 	& 0 (31.2)		& 0 (98.1) 		& 0 (90.0) 		& 0 (95.3) 		& 0 (97.0) 		& 0 (31.1)\bigstrut\\
Unphys. 	& 68.6 (68.8) & 1.8 (1.9) 	& 9.9 (10.0) 	& 4.5 (4.7) 	& 2.9 (3.0) 	& 68.5 (68.9)  \bigstrut\\
\hline
&\multicolumn{6}{c}{$X=0.3$} \bigstrut\\
\hline
Sol.		& 7.5 (0) 		& 84.7 (0) 		& 63.3 (0) 		& 69.8 (0)		& 78.0 (0) 		& 11.0 (0)\bigstrut\\
Causality 	& 0 (7.4) 		& 0 (84.5) 		& 0 (63.0) 		& 0 (69.4) 		& 0 (77.8) 		& 0 (10.8)  \bigstrut\\
Unphys. 	& 92.5 (92.6)	& 15.3 (15.5) 	&  36.7 (37.0) 	& 30.2 (30.6) 	&22.0 (22.2) 	& 89.0 (89.2) \bigstrut\\
\hline
&\multicolumn{6}{c}{$X=0.7$} \bigstrut\\
\hline
Sol.		& $\le 0.2$ 	& 29.0 (0) 		& 13.3 (0) 		& 11.1 (0) 		& 17.8 (0) 		& $\le 0.5$\bigstrut\\
Causality 	& $\le 0.2$	& 0 (28.8) 		& 0 (13.2) 		& 0 (10.9) 		&  0 (17.5)  	& $\le 0.5$  \bigstrut\\
Unphys. 	& $\ge 99.6$	& 71.0 (71.2) 	& 86.7 (86.8) 	& 88.9 (89.1) 	& 82.2 (82.5) 	& $\ge 99.0$ \bigstrut\\
\hline
\end{tabular}
\end{table*}

\subsection{Method 2: Bayesian Analysis}

We also investigate the effects of the chemical composition and the touchdown radius by conducting a Bayesian analysis. Our calculation method is similar to that of the previous study in \citet{Ozel_2016} but we include a new model parameter $h$ which determines the touchdown radius. Here we briefly describe the Bayesian method that we used. We refer to \citet{Ozel_2016} for more details of the Bayesian analysis used for the mass and radius estimation. 

Our Bayesian analysis was done on two measured quantities, apparent angular size ($A$) and touchdown flux ($F_{\rm TD,\infty}$), with several model parameters, radius ($R$) and mass ($M$) of a neutron star, distance ($D$), spin frequency ($f_{\rm NS}$), color correction factor ($f_{\rm c}$), hydrogen mass fraction of the photosphere ($X$), and $h$ for the touchdown radius ($h=2R/\rph$).  
From the Bayes' theorem, the posterior probability distribution $P(\vec{\theta}| {\rm data})$ of the parameter set $\vec{\theta}=\{ R, M, D,  f_{\rm NS}, f_{\rm c}, X, h \} $ on the given measured data ($F_{\rm TD,\infty}$, $A$) is expressed as 
\be
P(\vec{\theta} | {\rm data}) = \frac{P({\rm data} | \vec{\theta} ) P(\vec{\theta})}{P({\rm data})}~,
\ee
where $P({\rm data}|\vec{\theta})$ is the likelihood and $P({\rm data})$ is the evidence. $P(\vec{\theta})$ is the prior of the parameter set $\vec{\theta}$ of the model, which is composed of the multiplication of the prior distributions of the parameters as follows
\be
P(\vec{\theta}) = P(R) P(M) P(D) P(f_{\rm NS}) P(f_{\rm c}) P(X) P(h)~.
\ee

We assume that the prior distributions of the radius and the mass of a neutron star, $P(R)$ and $P(M)$, are flat over given ranges, $[2.0,\, 20.0]$~km and $[0.65,\, 3.5]~\msun$, respectively. The distance ($D$) to individual source is measured independently so that the prior distribution of the distance $P(D)$ is a normal distribution taken from the measurement except for EXO~1745--248 and KS~1731--260 for which a flat prior distribution is used (see Table~\ref{ta01}). The prior distribution of the spin frequency of the neutron star $P(f_{\rm NS})$ is assumed to be flat between 250~Hz and 650 Hz if a measured $f_{\rm NS}$ is not available. For the source which has a measured value of $f_{\rm NS}$,  $P(f_{\rm NS})$ is chosen as a delta function at the given value. We assume that $P(f_{\rm c})$ is flat over $[1.35,\, 1.45]$. Unlike our MC sampling, we do not fix $X$ nor $h$ for any source in our Bayesian analysis. Rather we use flat prior distributions for these two model parameters and then obtain their posterior distributions from which we could investigate their effects. Furthermore, based upon the posterior distributions, we can investigate whether there exists any statistical trend for these two model parameters including a correlation between them. The Bayesian approach is complementary to the MC sampling and provides a different aspect to study the effects of the chemical composition of the photosphere and the touchdown radius. We assume that $P(X)$ and $P(h)$ are flat over $[0.0,\,0.7]$ and $[0.0,\,2.0]$, respectively. 
The posterior distributions of the model parameters were obtained by conducting the Markov-Chain Monte Carlo (MCMC) simulations. We generated $2\times10^6$ MCMC samples for each case under investigation based on the Metropolis-Hastings algorithm.

\section{Results} \label{s03}

We apply our MC sampling and Bayesian analysis to six LMXBs that show PRE XRBs. Table \ref{ta01} presents the observed values with measurement uncertainties for these six sources. 
We choose and use the same observational values in \citet{Ozel_2016} who analyzed these sources to estimate mass and radius by conducting a Bayesian analysis. But they did not consider the possibility that touchdown occurs at a larger radius than the neutron star radius. A subset of these six sources (three sources) were analyzed by \citet{Steiner_2010} who investigated the effect of the touchdown radius by using the MC sampling. But they did not consider the correction terms on the touchdown flux and the apparent angular size. We note that the observational values in \citet{Steiner_2010} are slightly different from those in Table \ref{ta01} which are identical to those in \citet{Ozel_2016}. 
Since the aim of the current work is to investigate the effects of the photospheric composition and the touchdown radius, we focus on the general trend that is commonly seen in six sources with varying $X$ and $h$ instead on the results of individual sources. 
We refer to \citet{Steiner_2010,Ozel_2016} for more details of individual sources including the general information on each source and the mass and radius estimation carried out with different methods.

\subsection{Results from MC Sampling}

The results obtained with the MC sampling (including our iterative method) for the six sources are shown in Tables ~\ref{table:statistics1} to \ref{table:statistics3} and Figures \ref{fig:4U1820-30} to \ref{fig:4U1608-52}. 
Tables~\ref{table:statistics1} to \ref{table:statistics3} show the statistics of our MC sampling for the six sources. Out of $10^6$ samples, the resulting events fall into one of three cases. The numbers in Tables~\ref{table:statistics1} to \ref{table:statistics3} are the percentile of each case. The "solution" corresponds to the case that the physically allowed solution of mass and radius, either $\beta_-$ or $\beta_+$, was found. Since one sampling for $\alpha^*$ and $\gamma^*$ could result in two solutions ($\beta_\pm$, see Figure \ref{fig:beta_solution}), we count each solution separately. Not all physically allowed solutions are true solutions due to the causality limit ($\beta < 1/2.94$). The "causality" corresponds to this case, i.e., the events in the "solution" that violate the causality limit.   
As explained earlier in Section \ref{M_MC}, the "unphysical" corresponds to the case that we can not find any physically meaningful solution by using the $\alpha$-criterion. These unphysical solutions give non-real values of mass or radius.  
The $M$-$R$ distributions in Figures \ref{fig:4U1820-30} to \ref{fig:4U1608-52} are plotted only with the "solution" events. As a result, the distribution in each panel even within the same source is drawn from different number of events. In each panel, the maximum probability (indicated with red color) is calculated with the number of "solution" events in Tables \ref{table:statistics1} to \ref{table:statistics3} and the minimum probability (indicated with blue color) is chosen as 1/1000 of the maximum probability. 
 
Since we fix both $X$ and $h$, the effect of each can be seen individually. 
The effect of $X$ on the mass--radius estimation is seen as follows for all of the six sources. As $X$ increases, mass/radius increases/decreases regardless of $h$. It seems that this effect of $X$ on mass and radius is not affected significantly by $h$. In other words, the change in the probability distributions of mass and radius (particularly the most probable values of mass and radius) with $X$ looks similar for all three values of $h$. Similarly, the effect of $h$ on the mass--radius estimation does not change much with $X$, either. Regardless of $X$, as $h$ decreases (i.e., touchdown occurs far from the neutron star surface), one of the two physically allowed solutions approaches close to and eventually crosses over the causality limit. 

The effect of $X$ on the mass--radius estimation can be explained based upon equations in Section \ref{M_MC}. Since  the effects of the correction terms are not large, $\alpha^* \approx \alpha$ and $\gamma^* \approx \gamma$. Then, $R \sim \sqrt{1-2\beta(X)}$ from equation (\ref{Eq_R}) because $\alpha \gamma$ is now independent of $X$ (or $\kappa$) from equations (\ref{alpha_star}) and (\ref{gamma_star}) with correction terms ignored. 
According to equation (\ref{alpha_star}), $\alpha$ increases with $X$.  
Figure \ref{fig:beta_solution} shows how $\beta_\pm$ changes as a function of $\alpha$.  
Since $\beta_-$ increases with $\alpha$, $R$ decreases with $X$. Since $M \sim R \beta = \beta \sqrt{1-2\beta}$, $M$ increases with $X$ for the $\beta_-$ solution. However, on the contrary to $\beta_-$, $\beta_+$ decreases with $\alpha$, which must result in the opposite trend, i.e., $R$/$M$ increases/decreases with $X$. Our results that correspond to the $\beta_+$ solution seem to show this trend. For example, the most probable values of mass and radius obtained with $h=2$ (i.e., $\rph=R$) follow this trend (see the distributions close to the causality limit in the upper three panels in Figures \ref{fig:4U1820-30} to \ref{fig:4U1608-52}) although the trend does not appear clearly because the distributions are cut off at the causality limit. Another example may be found in the case of $h=1$ (i.e., $\rph=2R$) in which the distributions corresponding to the $\beta_+$ solution seem to appear for large $X$ (see the distributions close to the causality limit in the middle three panels in Figures \ref{fig:4U1820-30} to \ref{fig:4U1608-52}) because the distributions for the $\beta_+$ solution tend to move in the lower-right direction ($\searrow$) in the $M$-$R$ space as $X$ increases and can cross over the causality limit. However, in case of $h=0$ (i.e., $\rph >> R$), we can not see any distribution corresponding to the $\beta_+$ solution because there is no physically allowed solution for $\beta_+$ (see Figure \ref{fig:beta_solution}). 

The explanation above based upon the solution behaviors (as shown in Figure \ref{fig:beta_solution}) also applies to understanding the effect of $h$. The distributions for mass and radius change with $h$ simply because the value of $h$ determines the behaviors of the $\beta_\pm$ solution. The distributions corresponding to the $\beta_+$ solution are more likely to appear with a larger value of $h$ which corresponds to the case that touchdown occurs closer to the neutron star surface. 

We have to mention that the individual effect of $X$ and $h$ that we find by fixing both of them was already investigated in previous studies. In \citet{Steiner_2010} who introduced $h$ in their MC sampling analysis, the same effect of $h$ on the mass-radius estimation was seen in their results although they did not consider the correction terms. But they did not investigate the effect of $X$ because they did not vary $X$ systematically. We note that among three sources that they analyzed, they fixed $X=0$ for 4U~1820--30 considering observational constraints on $X$ for this source, but did not fix $X$ for the other two sources (EXO 1745--248 and 4U~1608--52). The effect of $X$ was tested for some individual sources which were previously analyzed with the probability transformation method. For example, for EXO~1745--248, \citet{O_zel_2009} found that $X=0$ is favored after testing that $X \gtrsim 0.1$ does not result in any feasible solution with their method although they did not consider the correction terms to draw the conclusion for the effect of $X$. Thus, we think that our results obtained not only by including the correction terms and but also by using the MC sampling can show a systematic effect of $X$ (as well as of $h$) on the mass-radius estimation although their effects can be inferred from the governing equations. Since we include the correction terms in our analysis, our results also confirm that the effects of the correction terms are not large enough to qualitatively change the effects of $X$ and $h$. However, different values of fixed $X$ and $h$ result in different values of mass and radius. Figure \ref{fig:most_mr} shows the estimated masses and radii that we obtained with fixed values of $X$ and $h$ in comparison with those in \citet{Ozel_2016}. 
From Figure \ref{fig:most_mr}, one can conclude that the upper limit of the most likely radii for these six sources is $\sim 12.5$~km, regardless of the touchdown radius and the photospheric composition. Note that 
this upper bound is consistent with the bounds on the radii of neutron stars estimated from 
the tidal deformability of GW170817, which was measured by the LIGO and Virgo collaboration~\citep{Abbott:2018exr}.

Although the way that we treat $X$ is different from that of \citet{Steiner_2010}, it is worth comparing how the acceptance rate (i.e., the percentiles of the physically allowed solutions in the results of the MC sampling) changes with the correction terms included. For all of the common three sources (4U~1820--30, EXO 1745--248, and 4U~1608--52), we find that the acceptance rate increases as $h$ decreases, which is consistent with what \citet{Steiner_2010} found. Again this confirms that the correction terms do not significantly affect the overall solution behaviors of the governing equations that include the correction terms. We note that except for the $h=0$ case of EXO 1745--248, $X=0.7$ is strongly disfavored regardless of $h$ for all the three sources. In fact, in all of the six sources, we find the same trend of the acceptance rate such that the acceptance rate increases as $h$ decreases. 
As for the acceptance rate as a function of $X$, which could not be addressed in \citet{Steiner_2010}, we find that for all of the six sources, the acceptance rate decreases as $X$ increases. The trend of acceptance rate as a function of $h$ and $X$ is consistent with the results of our Bayesian analysis which are presented in the next section.   

\begin{table*}
\centering
\caption{Result of the Pearson correlation test between $X$ and $h$ over the posterior samples obtained within the 68\% credible region. \label{ta-pearsonr}}
\begin{tabular}{cccccc}
\hline \hline
\multicolumn{6}{c}{Pearson correlation coefficient (R)} \bigstrut\\
\hline
 4U 1820--30 &  SAX J1748.9--2021 & EXO 1745--248 &  KS 1731--260 & 4U 1724--207 & 4U 1608--52 \bigstrut\\
 -0.356 & -0.622 & -0.526 &-0.539 & -0.631 & -0.529 \bigstrut\\
\hline
\end{tabular}
\end{table*}

\subsection{Results from Bayesian Analysis}

The results obtained with our Bayesian analysis are shown in Figures \ref{fig-corner-1} to \ref{fig-corner-6}. 
In these figures, we select to show the results only for four relevant model parameters, $R$, $M$, $X$, and $h$, among the total seven for all of which we obtained the posterior distributions. We find that the posterior distributions of radius and mass for the six sources obtained from our Bayesian analysis are generally consistent with those from \citet{Ozel_2016}. The shape of distribution and location of most probable value of $(M,~R)$ seems to be similar (but are not identical) in both results even if we include an additional model parameter $h$.
From this comparison, one may conclude that the effect of $h$ is not large in determining $M$ and $R$ in the Bayesian analysis. 
However, more detailed comparison based upon the posterior distributions of $X$ and $h$ reveal new findings. First of all, the hydrogen mass fraction is likely to be small in the posterior distributions for all of the six sources. In our Bayesian analysis, the prior distributions of $X$ are uniform in $[0,~0.7]$ in all of the six sources, but the posterior distributions of $X$ are skewed to lower values and their medians are smaller than 0.35 (the mean of the prior distribution) in all of the six sources. The small values of $X$ favored in the posterior distributions are consistent with the acceptance rate decreasing with $X$ which was found from our MC sampling. These results on $X$ could imply that the photospheric composition is likely to be H-poor.

We pay attention to 4U~1820--30 for which the posterior distribution of $X$ is much smaller than those for the other sources. Small $X$ for this target is consistent with the observational constraint on this source, i.e., this target being an ultracompact binary with an H-poor donor \citep{1986Natur.323..105K,1987ApJ...315L..49S} and the theoretical constraint favoring the H-poor fuel \citep{2003ApJ...595.1077C}, both of which were already mentioned in \citet{Steiner_2010}. For this reason, both \citet{Steiner_2010} and \citet{Ozel_2016} fixed $X=0$ in their MC sampling and Bayesian analysis, respectively. We note that we choose a larger distance $8.4 \pm 0.6$~kpc for 4U~1820--30 in our Bayesian analysis between two available distance measurements in Table \ref{ta01} while \citet{Ozel_2016} used both distance measurements by combining them as a double Gaussian distribution. The shape of the mass-radius distribution obtained from our analysis for 4U~1820--30 does not look very different from that in \citet{Ozel_2016}, but our analysis has a larger value for most probable value of mass ($M \approx 1.99~\msun$ similar to the median) which may come from the choice of different distance. We note that among the six sources, 4U~1820--30 is the only source for which \citet{Ozel_2016} fixed $X=0$. 

Our analysis also reveals that $h$ is not highly populated around 2.0 (which corresponds to the case that touchdown occurs on the neutron star surface) in the posterior distribution of $h$. We find this behavior of $h$ in all of the six sources as well. The medians of $h$ in all of the six sources are smaller than $h=1$ which correspond to $\rph = 2 R$. In particular, as in the case of $X$, $h$ favors smaller values for 4U~1820--30 than for the other sources. Our results for such behavior of $h$ are consistent with the acceptance rate decreasing with $h$ which was found from our MC sampling. 
These results on $h$ could imply that touchdown is not likely to occur on the neutron star surface.

We also investigate the correlation between $X$ and $h$ based upon their posterior distributions   
by using the Pearson correlation test (Table \ref{ta-pearsonr}).
We find a relatively significant correlation between these two model parameters in five out of six sources for which $|R|>0.5$. The correlation looks weak for 4U 1820--30 in which $|R|<0.5$. Again, 4U~1820--30 shows an unusual feature in comparison with the other five sources. We note that the posterior distributions of both $X$ and $h$ for 4U~1820--30 are skewed more toward lower values.    
The (weak) anti-correlation between $X$ and $h$ could be explained based upon the equations in Section \ref{M_MC}. By using again the fact that the effects of the correction terms are not significant, equation (\ref{edd2}) leads to $(1+X) \sim \sqrt{1-\beta h}$, where we assume that $M$ and $\beta$ does not vary much with $X$ and $h$. As $X$ and $h$ vary within a certain range while the assumption on $M$ and $\beta$ is still valid, we can expect the anti-correlation between them, i.e., $h$ increases/decreases as $X$ decreases/increases. However, this anti-correlation may not be that strong when the assumption on $M$ and $\beta$ breaks down or the solutions for mass and radius can be found only for small values of $X$ or $h$ which might be the case for 4U 1820--30.      

We propose that the (weak) anti-correlation between $X$ and $h$ could be explained physically as well with the  relationship between the photospheric composition and the radiation pressure (which is determined by the Eddington flux as a function of the photospheric composition). 
If the photosphere is H-poor (i.e., small $X$), the radiation pressure decreases due to the small opacity with the same luminosity which is determined by the thermonuclear reactions occurring below the photosphere. The decreased radiation pressure does not push the photosphere to a large distance so touchdown is also likely to occur close to the neutron star surface, i.e., large $h$ or $h \approx 2$. In contrast, if the photosphere is H-rich, the photosphere can expand farther from the neutron star surface due to the large radiation pressure with large opacity and touchdown is likely to occur away from the neutron star surface. We note that this explanation for the weak anti-correlation between $X$ and $h$ is based upon the assumption that the fuel composition for thermonuclear burning which determines the luminosity is not correlated with the photospheric composition.
Finally, we point out that the most probable radii estimated with the Bayesian analysis, which includes the effects of both the touchdown radius and the photospheric composition, are less than 12.5~km, which is consistent with the results obtained with the MC sampling. This implies that the upper limit on the radii of the neutrons stars in six LMXBs analyzed in our study is valid, regardless of the statistical methods, when the uncertainties in the touchdown radius and the photospheric composition are taken into account.

 \section{Conclusion}\label{s04}

In order to investigate the effects of the chemical composition of the photosphere and the touchdown radius on the estimation of mass and radius of a neutron in LMXB that shows PRE XRB, we carry out statistical analyses on six LMXBs by using both a MC sampling and a Bayesian analysis. Following the idea of \citet{Steiner_2010}, we allow touchdown to occur away from the neutron star surface and introduce a new parameter $h$ in our statistical analyses. 
For the six LMXBs that we apply our MC sampling and Bayesian analysis to, we use the same observational data of \citet{Ozel_2016} who estimated the masses and radii of these six sources with their Bayesian analysis without $h$. 
In both methods, we solve the Eddington flux equation and the apparent angular area equation both of which include the correction terms. 

In our MC sampling, we fix both $X$ (hydrogen mass fraction of the photosphere) and $h$ at three values and investigate how the mass and radius estimation changes with these values of $X$ and $h$. 
We confirm that the mass and the radius estimated with our MC sampling change with $X$ and $h$ as expected from the analytic solutions of the equations without correction terms. 
This result implies that the effects of the correction terms are not large enough to qualitatively change the effects of $X$ and $h$ although different values of fixed $X$ and $h$ result in different values of mass and radius. We find that the acceptance rate which is the ratio of the physically allowed solutions to the total realizations of our MC sampling decreases with $X$ and $h$ in all of the six sources. This implies that small values of $X$ and $h$ are favored statistically, i.e. the photosphere is likely to be H-poor regardless of the energy--generation mechanism below the photosphere and touchdown is likely to occur away from the neutron star surface. But we can not conclude any correlation between these two parameters based upon the MC sampling.

As a complementary analysis to the MC sampling, we choose flat prior distributions for both $X$ and $h$ in our Bayesian analysis instead of fixing them. 
We find that the posterior distributions of $X$ and $h$ are consistent with the acceptance rate decreasing with $X$ and $h$ which was found from our MC sampling. In all of the six sources, the posterior distributions of $X$ and $h$ favor small values. 
By taking advantage of the Bayesian analysis, we investigate the correlation between $X$ and $h$ based upon the selected samples within the $68\%$ confidence regions of the posterior distributions. We find that $X$ and $h$ are (weakly) anti-correlated in all of the six sources, which could be qualitatively understood with the Eddington flux equation. 
We propose that the anti-correlation between $X$ and $h$ could be explained physically as well with the relationship between the radiation pressure and the opacity. 
Since the radiation pressure increases with opacity (or $X$) with the same luminosity on the photosphere,  
it is likely that touchdown occurs farther away from the neutron star surface (small $h$) after the strong radiation pressure (with large $X$) pushes the photosphere to a larger distance.  
Since we find this correlation by analyzing six sources, it is worth investigating further whether the same correlation can be found for other LMXBs that show PRE XRBs. 
We find that the upper bound of the most probable radii of the neutron stars in the six LMXBs analyzed with both the MC sampling and the Bayesian analysis in the current study is around 12.5 km when the uncertainties in the touchdown radius and the photospheric composition are taken into account. It is interesting that this upper bound is similar to that placed by the LIGO and Virgo observations with the measurement of the tidal deformability of GW170817. Thus it will be worth investigating whether a similar upper bound can be put on the radii of neutron stars in other LMXBs.


\begin{acknowledgements}
We thank the referee for suggesting us to consider the effect of the touchdown radius together with that of the chemical composition of the photosphere. The referee's suggestion and comments have greatly improved our earlier work.    
This work was supported by 
the National Research Foundation of Korea (NRF) grant funded 
by the Korea government (Ministry of Science and ICT) (No. 2016R1A5A1013277). 
MK and CHL were also supported by NRF grant funded by the Korea government (Ministry of Education)  (No. 2018R1D1A1B07048599).
YMK was also supported by NRF grants funded by the Korea government (MSIT) (No. 2019R1C1C1010571).
KHS was also supported by NRF grant funded by Ministry of Education (NRF-2015H1A2A1031629-Global Ph.D. Fellowship Program).
KK was also supported by NRF grant funded by Ministry of Education (No. 2016R1D1A1B03936169). 
We acknowledge the hospitality at the APCTP where part of this work was done.
MK and YMK contributed equally to this work as co-first authors by carrying out the Monte Carlo sampling and the Bayesian analysis, respectively. 
\end{acknowledgements}

\bibliographystyle{aa}
\bibliography{biblio}

\end{document}